%% file: main.tex
\documentclass[fleqn,usenatbib]{mnras} 

\usepackage{newtxtext,newtxmath}

\usepackage[T1]{fontenc}
\usepackage{ae,aecompl}

\usepackage{array}

\newcolumntype{?}{!{\vrule width 1.5pt}}
\usepackage{afterpage}

\usepackage{multirow}
\usepackage{graphicx}	
\usepackage{amsmath}	
\usepackage{amssymb}	
\usepackage[usenames]{color}
\usepackage{bm}		
\usepackage{booktabs,makecell}
\usepackage{adjustbox}

\input{addresses.tex}

\newcommand{\redmapper}{\textit{redMaPPer}}
\newcommand{\redmagic}{\textit{redMaGiC}}

\newcommand{\Nz}{$n(z)$}
\newcommand{\Nuz}{$n_{{\rm u}}(z)$}
\newcommand{\Nrz}{$n_{{\rm r}}(z)$}
\newcommand{\phiu}{n_{{\rm u}}}
\newcommand{\phir}{n_{{\rm r}}}

\newcommand{\phipz}{n_{{\rm pz}}}
\newcommand{\phiwz}{n_{{\rm wz}}}
\newcommand{\phidel}{n_{\Delta}}

\newcommand{\rmin}{r_{{\rm min}}}
\newcommand{\rmax}{r_{{\rm max}}}

\newcommand{\wur}{\bar w_{\rm{ur}}}

\newcommand{\zmax}{z_{\rm{max}}}

\newcommand{\avg}[1]{\langle #1 \rangle}

\definecolor{purple}{RGB}{150,0,200}

\title[Cross-Correlation Redshifts in DES: Methods and Systematics]{Dark Energy Survey Year 1 Results:  Cross-Correlation Redshifts - Methods and Systematics Characterization}

\makeatletter
\def \blfootnote{\xdef\@thefnmark{}\@footnotetext}
\makeatother

\input{authors_MNRAS.tex}		
\date{\today}

\begin{document}     
\label{firstpage} 
\pagerange{\pageref{firstpage}--\pageref{lastpage}}%
\maketitle        	 

\begin{abstract}
We use numerical simulations to characterize the performance of a clustering-based method to calibrate photometric redshift biases. In particular, we cross-correlate the weak lensing (WL) source galaxies from the Dark Energy Survey Year 1 (DES Y1) sample with \redmagic\ galaxies (luminous red galaxies with secure photometric redshifts) to estimate the redshift distribution of the former sample. The recovered redshift distributions are used to calibrate the photometric redshift bias of standard photo-$z$ methods applied to the same source galaxy sample. We apply the method to three photo-$z$ codes run in our simulated data: Bayesian Photometric Redshift (BPZ), Directional Neighborhood Fitting (DNF), and Random Forest-based photo-$z$ (RF).  We characterize the systematic uncertainties of our calibration procedure, and find that these systematic uncertainties dominate our error budget. The dominant systematics are due to our assumption of unevolving bias and clustering across each redshift bin, and to differences between the shapes of the redshift distributions derived by clustering vs photo-$z$'s. The systematic uncertainty in the mean redshift bias of the source galaxy sample is $\Delta z \lesssim 0.02$, though the precise value depends on the redshift bin under consideration. We discuss possible ways to mitigate the impact of our dominant systematics in future analyses.
\end{abstract}
\blfootnote{$^{\star}$ E-mail: mgatti@ifae.es}
\blfootnote{$^{\dag}$ E-mail: pvielzeuf@ifae.es}
\blfootnote{$^{\ddag}$ Einstein fellow}
\blfootnote{ Affiliations are listed at the end of the paper.}
\setcounter{footnote}{1}




\input{Introduction}
\input{Methodology}

\input{Data}
\input{Systematics_old}

\input{Discussion}
\input{Conclusion}
\input{Aknowledgements}



\bibliographystyle{mnras}		  
\bibliography{bibliography,Y1bib}



\appendix
\input{Appendix}
\section*{Affiliations}
\input{affiliation1.tex}


\bsp	
\label{lastpage}
\end{document}

%% file: authors_MNRAS.tex
\author[M. Gatti, P. Vielzeuf et al.]{
\parbox{\textwidth}{
\Large 
M.~Gatti$^{1\star}$,
P.~Vielzeuf$^{1\dag}$,
C.~Davis$^{2}$,
R.~Cawthon$^{3}$,
M.~M.~Rau$^{4}$,
J.~DeRose$^{5,2}$,
J.~De Vicente$^{6}$,
A.~Alarcon$^{7}$,
E.~Rozo$^{8}$,
E.~Gaztanaga$^{7}$
% ...
}
}
\vspace{0.2cm}

%% file: Introduction.tex
\section{Introduction}
Current and future large photometric galaxy surveys like DES\footnote{https://www.darkenergysurvey.org/} \citep{DES}, KiDS\footnote{http://kids.strw.leidenuniv.nl/} \citep{KIDS}, HSC\footnote{https://subarutelescope.org/Projects/HSC/} \citep{HSC}, LSST\footnote{https://www.lsst.org/} \citep{LSST}, Euclid\footnote{http://sci.esa.int/euclid/} \citep{EUCLID}, and WFIRST\footnote{https://wfirst.gsfc.nasa.gov/} \citep{WFIRST} will map large volumes of the Universe, measuring the angular positions and shapes of hundreds of millions (or billions) of galaxies. This will allow cosmological measurements with an unprecedented level of precision, leading to a considerable step forward in our understanding of cosmology and particularly of the nature of dark energy. To capitalize on their statistical constraining power, these surveys require accurate characterization of the redshift distributions of selected galaxies, which presents a considerable challenge in the absence of complete spectroscopic coverage.

Given the large amount of forthcoming photometric data, obtaining a spectroscopic redshift for every individual source is unfeasible: spectroscopy of large samples is time-consuming and expensive, and it is usually restricted to the brightest objects of any given sample.  Because of this limitation, photometric surveys provide redshift estimates for each galaxy based on that galaxy's multi-band photometry, a technique called photometric redshift, or photo-$z$. There exists a large variety of photo-$z$ methods \citep[e.g.][]{Hildebrandt2010,Sanchez2014}. 
However, unrealistic SED templates,  degeneracies between colors and redshift, and unrepresentative spectroscopic samples for both training and calibration ultimately limit the performance of photo-$z$ methods \citep{Lima2008,Cunha2009,Newman2015,Bezanson2016,Masters2017}.

Clustering-based redshift estimation methods \citep{Newman2008,MattewsNewman2010,Menard2013,Schmidt2013} constitute an   interesting alternative to infer redshift distributions, since they are more general and do not suffer the above limitations. Briefly, one uses the fact that the correlation amplitude between a sample
with unknown redshifts and a reference sample with known redshifts in some narrow
redshift bin can be related to the fraction of galaxies in the
unknown sample that lie within the redshift range of the reference sample.

Clustering-based redshift estimators have been studied and applied both to simulations and to data \citep[e.g.][]{Menard2013,McQuinn2013,Schmidt2013,Scottez2016,Scottez2017,Hildebrandt2017,vanDaalen2017, DAVIS}. \cite{Hildebrandt2017} cross-correlated the source galaxies used in the KiDS cosmological analysis with galaxies from zCOSMOS \citep{Lilly2009} and DEEP2 \citep{Newman2013}.  Unfortunately, the small ($\leq 1\ \deg^2$) area covered by these reference galaxy samples severely limited the usefulness of the resulting cross-correlation analyses.  Consequently, \cite{Hildebrandt2017} ultimately chose to rely on traditional photo-$z$ methods in deriving the KiDS cosmological constraints. 

%

The DES Y1 cosmological analyses rely on a different strategy.  Instead of using a small spectroscopic sample as reference, we use red-sequence galaxies from the DES Y1 red-Sequence Matched-filter Galaxy Catalog \citep[\redmagic,][]{Rozo2016}.  The \redmagic\ algorithm is designed to select galaxies with high quality photometric redshift estimates.  While the reliance on \redmagic\ photometric redshifts may be a source of concern for the cross-correlation program, the vastly superior statistical power of the sample renders the resulting cross-correlation constraints competitive with traditional photo-$z$ methods. 

The DES Y1 analysis attempts to combine traditional photo-$z$ methods with cross-correlation techniques.  In particular, motivated by the fact that the DES Y1 cosmological analyses are primarily sensitive to an overall redshift bias in the photometric redshift estimates \citep{photoz,shearcorr,keypaper}, we have sought to use cross-correlation methods to verify and calibrate the redshift bias of traditional photo-$z$ methods.  By combining these two techniques we benefit from the strength of both methods, while ameliorating their respective weaknesses. This calibration strategy is fully implemented in the DES Y1 cosmic shear and combined two-point function analysis \citep{methodpaper,shearcorr,keypaper}. 

This paper characterizes the performance and systematic uncertainties of our method for calibrating photometric redshift biases in the DES Y1 source galaxy sample via cross-correlation with \redmagic\ galaxies. Specifically, we implement our method on simulated data, introducing sources of systematic uncertainty one at a time to arrive at a quantitative characterization of the reliability and accuracy of our method.  A companion paper  \citep{xcorr} implements the photometric calibration method developed here to enable DES Y1 cosmology analyses, while a second companion paper \citep{redmagicpz} uses cross-correlations to validate the photometric redshift performance of the \redmagic\ galaxies themselves.

The paper is organized as follows. In $\S$\ref{methods_general} we present the methodology we use to calibrate photo-$z$ posteriors using clustering-based redshift estimation. The simulations and the samples used are described in $\S$\ref{datasets}. $\S$\ref{systematics} is devoted to the study and quantification of the systematic error of our method.  In $\S$\ref{discussion} we further discuss some aspects of clustering-based redshift estimation techniques and how our method could be improved upon in the future. Finally we present our conclusions in $\S$\ref{conclusions}.

%% file: Methodology.tex
\section{Methodology}\label{methods_general}

In DES Y1 we will use clustering-based redshift estimates to correct the photo-$z$ posterior distributions of a given science sample. We defer the description (and the binning) of the particular samples (unknown and reference) adopted in this work to $\S$\ref{datasets}, while keeping the description of the methodology as general as possible. Here, ``unknown'' always refers to the photometric galaxy sample for which we wish to calibrate photometric redshift biases, while ``reference'' refers to the galaxy sample with known, highly accurate redshifts (be they spectroscopic or photometric).   

Our methodology divides into two steps:
\begin{enumerate}
\item We first estimate the redshift distribution of the unknown galaxy sample by cross-correlating with the reference sample. Note that the reference sample does not necessarily have to span the full redshift interval of the unknown sample.
\item We then use the recovered redshift distribution to calibrate the redshift bias of the unknown galaxy population by finding the shift $\Delta z$ which brings the photo-$z$ posterior in better agreement with the redshift distribution obtained through cross-correlations.
\end{enumerate}

\subsection{First step: clustering-based redshift estimates} \label{method_clustering-z}

In the literature a variety of methods to recover redshift distributions based on cross-correlation have been discussed \citep{Newman2008,Menard2010,Schmidt2013,McQuinn2013}. The underlying idea  shared by all methods is that the spatial cross-correlation between two samples of objects is non-zero only in case of 3D overlap. Let us now consider two galaxy samples:

\begin{enumerate}
  \item    An \textit{unknown} sample, whose redshift distribution \Nuz\  has to be recovered. 
  \item   A \textit{reference} sample, whose redshift distribution \Nrz\ is known (either from spectroscopic redshifts or from high-precision photometric redshifts). The reference sample is divided into narrow redshift bins. 
 \end{enumerate}

To calibrate the redshift distribution of the unknown sample we bin the reference sample in narrow redshift bins, and then compute the angular cross-correlation signal $w_{ur}$ between the unknown sample and each of these reference redshift bins.   Under the assumption of linear biasing, we find

\begin{equation}
w_{{\rm ur}}(\theta) = \int  dz'\ \phiu(z')\phir(z')
 	b_u(z')b_r(z')w_{DM}(\theta,z') ,
	\label{crosscorr}
\end{equation}
where $\phiu(z')$ and $\phir(z')$ are the unknown and reference sample redshift distributions (normalized to unity over the full redshift interval), $b_u(z')$ and  $b_r(z')$  are the biases of the two samples, and $w_{DM}(\theta,z')$ is the dark matter 2-point correlation function. 

In this paper we implement three different clustering-based methods: Schmidt's method \citep{Schmidt2013}, M\'enard's method \citep{Menard2013,Scottez2016}, and Newman's method \citep{Newman2008,MattewsNewman2010}. We briefly describe each of the three methods. A comparison of the three methods is presented in $\S$\ref{subsect:scale_method}.  At the end, we have opted for using Schmidt's method for our fiducial analysis.

\textbf{Schmidt's method}: In \cite{Schmidt2013}, the authors use a ``1-angular bin'' estimate of the cross-correlation signal. This is achieved by computing the number of sources of the unknown sample in a physical annulus around each individual object of the reference sample, from a minimum comoving distance $\rmin$ to a maximum distance $\rmax$. Our fiducial choice for the scales is from 500 kpc to 1500 kpc \footnote{Even though these scales are clearly non-linear, these non-linearities do not have a significant impact on the methodology, as demonstrated in \cite{Schmidt2013} and in this paper. See $\S$\ref{subsect:scale_method} for a discussion concerning the choice of scales.}. In addition, each object of the unknown sample is weighted by the inverse of the distance from the reference object, which has been shown to increase the S/N ratio of the measurement \citep{Schmidt2013}. We use the Davis \& Peebles \citep{DavisPeebles1983} estimator of the cross-correlation signal,
\begin{equation}
\label{DavisPeeblesestimator}
\wur =\frac{N_{Rr}}{N_{Dr}}\frac{\int_{r_{min}}^{r_{max}}dr' W(r')\left[D_uD_r(r')\right]}{\int_{r_{min}}^{r_{max}}dr' W(r')\left[D_uR_r(r')\right]}-1,
\end{equation}
where $D_uD_r(r')$ and $D_uR_r(r')$ are respectively data--data and data--random pairs, and $W(r') \propto 1/r'$ the weight function. The pairs are properly normalized through ${N_{Dr}}$ and ${N_{Rr}}$, corresponding to the total number of galaxies in the reference sample and in the reference random catalog. 

The  Davis \& Peebles estimator is less immune to window function contamination than the Landy \& Szalay estimator \citep{LandySzalay1993}, since it involves using a catalog of random points for just one of the two samples. We choose to use the  Davis \& Peebles estimator so as to avoid creating high-fidelity random catalogs for the DES Y1 source galaxy sample (our unknown sample): the selection function depends on local seeing and imaging depth, resulting in a complex spatial selection function. We therefore decide to use a catalog of random points only for the reference sample, whose selection function and mask are well understood \citep{wthetapaper}. 

Assuming the reference sample is divided into sufficiently narrow bins centered at $z$, we can approximate $ \phir(z') \propto N_r \delta_D (z-z')$ (with $\delta_D $ being Dirac's delta distribution, and $N_r$ being the number of galaxies in the reference bin) and  invert eq. (\ref{crosscorr}) to obtain the redshift distribution of the unknown sample: 
\begin{equation}
\label{menard2}
\phiu(z) \propto  \wur(z)  \frac{1}{b_u(z)} \frac{1}{b_r(z)}
  \frac{1}{\bar{w}_{DM}(z)},
\end{equation}
where barred quantities indicate they have been ``averaged'' over angular scales, reflecting the fact that we are using 1-angular bin estimates of the correlation while weighting pairs by their inverse separation. The proportionality constant is obtained from the requirement that $\phiu(z)$ has to be properly normalized.

In principle, the redshift evolution of the galaxy--matter biases and of $\bar{w}_{DM}(z)$ could be estimated by measuring the 1-bin autocorrelation function of both samples as a function of redshift:
\begin{flalign}
\label{autocorrelation1}
\bar{w}_{rr,z} = \int dz'   \left[b_r(z') n_{r,z}(z')\right]^2 \bar{w}_{DM}(z') ,\\
\label{autocorrelation2}
\bar{w}_{uu,z} =\int dz'  \left[b_u(z')n_{u,z}(z')\right]^2 \bar{w}_{DM}(z'),
\end{flalign}
where $n_{r,z}(z')$  and $n_{u,z}(z')$ are the redshift distributions of the reference and unknown samples binned into narrow bins centered in $z$. If the bins are sufficiently narrow so as to consider the biases and $\bar{w}_{DM}$ to be constant over the distributions, they can be pull out of the above integrals. Knowledge of the redshift distributions of the narrow bins is then required to use eqs. (\ref{autocorrelation1}) and (\ref{autocorrelation2}) to estimate $b_r$, $b_u$ and $\bar{w}_{DM}$.

In our fiducial analyses we do not attempt to correct for the redshift evolution of the galaxy--matter bias and of the dark matter density field. Rather, we assume $b_r$, $b_u$ and $\bar{w}_{DM}$ to be constant within each photo-z bin, and use the simulations to estimate the systematic error induced by this assumption. This choice is motivated and discussed in more details in $\S$\ref{subsect:biasev} and Appendix \ref{biascorrref}. Under this assumption, eq. (\ref{menard2}) reduces to:
\begin{equation}
\label{menard3}
\phiu(z) \propto  \wur(z),
\end{equation}
where the proportionality constant is again obtained requiring a proper normalization for $\phiu(z)$.

\textbf{M\'enard's method} \citep{Menard2013}: the method of M\'enard differs from that of Schmidt in how the 1-angular bin estimate of the cross-correlation signal is measured. In particular, for M\'enard's method the correlation function is measured as a function of angle, and the recovered correlation function is averaged over angular scales via
\begin{equation}
\wur(z)=\int_{\theta_{min}}^{\theta_{max}}d\theta\ W(\theta)w_{ur}(\theta,z) ,
	\label{crosscor2}
\end{equation}
where $W(\theta) \propto \theta^{-\gamma}$ is a weighting function. We assume $\gamma=1$ to increase the S/N. The integration limits in the integral in eq. (\ref{crosscor2})  correspond to fixed physical scales (500 kpc to 1500 kpc). As can be seen, the primary difference between M\'enard's method and Schmidt's method is whether one computes the angular correlation function first followed by a weighted integral over angular scales, or whether one performs a weighted integral of pairs first, and then computes the angular correlation function.  

\textbf{Newman's method} \citep{Newman2008,MattewsNewman2010}: this method assumes that all the correlation functions can be described by power laws $\xi(r)=(r/r_0)^{-\gamma}$.  Adopting a linear bias model, this allows one to relate the measured cross-correlation signal to \Nuz\  and to quantities computable from a given cosmological model.  Specifically, one has
\begin{equation}
\label{newman_eq}
w_{\rm ur}(\theta,z)=\frac{\phiu(z) H(\gamma_{\rm ur})r_{0,{\rm ur}}^{\gamma_{\rm ur}}\theta^{1-\gamma_{\rm ur}}D_{A}^{1-\gamma_{\rm ur}}}{d\chi/dz}.
\end{equation}
Here $\gamma_{\rm ur}$ corresponds to the power law slope of the correlation function, while $H(\gamma_{ur})=\Gamma(1/2)\Gamma((\gamma_{ur}-1)/2)/\Gamma(\gamma_{ur}/2)$. $D_{A}(z)$  and $\chi(z)$ are respectively the angular size distance and the comoving distance at a given redshift. 

We fit the observed cross-correlation signal using a function of the form $w_{\rm ur}(\theta,z)=A_{\rm ur}(z)\theta^{1-\gamma_{\rm ur}}-C_{\rm ur}$. With respect to Schmidt's and M\'enard's methods, we note that the Newman's implementation introduces two extra degrees of freedom ($\gamma_{\rm ur}$ and $C_{\rm ur}$). The index $\gamma_{\rm ur}$ is estimated from the arithmetic mean of the indexes of the unknown and reference autocorrelation functions (see below). The parameters $A_{\rm ur}(z)$, and $C_{\rm ur}$ are obtained through chi square minimization; we estimate the covariance needed for the fit through jackknife resampling. Setting our two expressions for $w_{\rm ur}$ equal to each other, and solving for the redshift distribution, we arrive at
\begin{equation}
\label{newman_eq_2}
\phiu(z)= \frac{d\chi/dz}{D_{A}^{1-\gamma_{\rm ur}}H(\gamma_{\rm ur})r_{0,ur}^{\gamma_{\rm ur}}}A_{\rm ur}(z).
\end{equation}
Under the assumption of linear bias, both the index of the cross-correlation function and its correlation length can be calculated from the unknown and reference autocorrelations.  One has $\gamma_{\rm ur}=(\gamma_{\rm uu}+\gamma_{\rm rr})/2$ and $r_{0,{\rm ur}}^{\gamma_{\rm ur}}= (r_{0,{\rm uu}}^{\gamma_{\rm uu}}r_{0,rr}^{\gamma_{\rm rr}})^{1/2}$. \footnote{We note that if we assume constant (scale-independent) bias, then $\gamma_{\rm uu}=\gamma_{\rm rr}$. Nonetheless, we compute $\gamma_{ur}$ as the arithmetic mean of $\gamma_{uu}$ and $\gamma_{rr}$ to follow Matthew\&Newman's original recipe.} A first guess value for $r_{0,uu}$ can be inserted in eq. (\ref{newman_eq_2}) to estimate the redshift distribution, which can be inserted back in eq. (\ref{newman_eq}) to refine the value of $r_{0,\rm uu}$. The whole procedure is repeated until convergence.   

\subsection{Second step: correcting photo-$z$ posterior} \label{correction_photoz}

Given an unknown galaxy sample, one can readily use photo-$z$ techniques to estimate the corresponding redshift distribution.  Here, we seek to use the redshift distribution recovered via cross-correlation to calibrate the photometric redshift bias of the photo-$z$ posterior.  We investigated two approaches:

\begin{itemize}
\item criteria ${\rm I}$ - shape matching. Let $\phipz(z)$ be the photo-$z$ posterior for the unknown galaxy sample and $\phiwz(z)$ the redshift distribution recovered via cross-correlations. The corrected photo-$z$ posterior is defined as $\phidel(z) \equiv \phipz(z-\Delta z)$, where $\Delta z$
is the photometric redshift bias. The photo-$z$ bias is calibrated matching the shapes of $\phidel(z)$ and $\phiwz(z)$ within the redshift interval covered by the reference sample.

\item criteria ${\rm II}$ - mean matching. Let $\avg{z}_{\Delta}$ be the mean of $\phidel(z)$ and $\avg{z}_{\rm wz}$ the mean of $\phiwz (z)$. The photo-$z$ bias is calibrated requiring $\avg{z}_{\Delta}$ and $\avg{z}_{\rm wz}$ to match. Note that the means have to be estimated over the same redshift range.
\end{itemize}

Quantitatively, the matching is done using a likelihood function to solve for the photometric redshift bias of the photo-$z$ posterior.  We recall that we do not attempt to debias higher order moments of the photo-$z$ posterior as the cosmological probes in the accompanying DES Y1 analysis are primarily sensitive to the mean of this distribution \citep{keypaper,photoz,methodpaper,shearcorr}. The log-likelihoods for the parameter $\Delta z$ for each of the two matching criteria are defined via
\begin{equation}
\label{likeA}
	\mathcal{L_{\rm I}} = -\frac{1}{2} \chi^2 \left( e^k \phidel ;\ \phiwz; \ \hat{\Sigma}^{-1}_{\mathrm{wz}} \right) \\ + \mathrm{Prior} \left(k, \Delta z \right),
\end{equation}
and
\begin{equation}
\label{likeB}
	\mathcal{L_{\rm II}} = -\frac{1}{2} \chi^2 \left( \avg{z}_{\Delta}; \ \avg{z}_{{\rm wz}} ; \ \hat{\Sigma}^{-1}_{\avg{z}_{{\rm wz}}} \right) \\ + \mathrm{Prior} \left( \Delta z \right) .
\end{equation}
Note these likelihoods can account for the existence of \it a priori \rm estimate of the photometric redshift bias $\Delta z$.  In the above equations, $e^k$ is a relative normalization factor that rescales $\phidel(z)$, which is properly normalized to unity over the full redshift interval, to a distribution that is normalized to unity over the range of $\phiwz(z)$.

The quantity $\hat \Sigma$ for each of the likelihoods is the appropriate covariance matrix from the cross-correlation analysis. They are estimated from simulated data through a jackknife (JK) approach, using the following expression \citep{Norberg2009}:
\begin{equation}
\hat{\Sigma}(x_i,x_j)=\frac{(N_{JK}-1)}{N_{JK}} \sum_{k=1}^{N_{JK}} (x_i^k-\bar{x_i})(x_j^k-\bar{x_j}),
\end{equation}
where the sample is divided into $N_{JK}=1000$ sub-regions of roughly equal area ($\sim 1$ deg$^2$),  $x_i$ is a measure of the statistic of interest in the i-th bin of the k-th sample, and $\bar{x_i}$ is the mean of our resamplings. The jackknife regions are safely larger than the maximum scale considered in our clustering analysis. The Hartlap correction \citep{Hartlap2007} is used to compute the inverse covariance.

Finally, despite the fact that our reference sample (\redmagic\ galaxies) spans the redshift interval $[0.15,0.85]$ in our simulations, in practice, in criteria ${\rm II}$ (mean matching), we restrict ourselves to a narrower redshift range, defined by the intersection of $[0.15,0.85]$ and $[\avg{z}_{\rm wz}-2\sigma_{\rm wz},\avg{z}_{\rm wz}+2\sigma_{\rm wz}]$, where $\sigma_{\rm wz}$ is the root mean square of the redshift distribution $\phiwz(z)$.  We have found that this choice increases the accuracy and robustness of our method by minimizing systematics (e.g. lensing magnification) associated with regions in which there is little intrinsic clustering signal.   $\S$\ref{subsect:interval choice} quantifies the impact of this choice on our results.  We do not shrink the interval used for matching under criteria ${\rm I}$ (shape matching), as this procedure is inherently less sensitive to noise and biases in the tails.

One important feature of our analysis is that, when treating multiple WL source redshift bins, we treat each bin independently.  In practice, there are clear statistical correlations between bins, as revealed by significant off-diagonal elements in the jackknife covariance matrix.  However, as we demonstrate below, our analysis is easily systematics dominated.  This has an important consequence, as we have found that attempting a simultaneous fit to all WL source redshift bins clearly produces incorrect results: systematic biases in one bin get propagated into a different bin via these off-diagonal terms, throwing off the best fit models for the ensemble.  By treating each  bin independently, we find that we can consistently recover numerically stable, accurate (though systematics dominated) estimates of the photometric redshift bias.\footnote{In principle, neglecting correlations between different bins should result in an underestimation of the statistical uncertainty. In practice, this effect is negligible.}

We sampled the likelihood in eqs. (\ref{likeA}) and (\ref{likeB}) using the affine-invariant Markov Chain Monte Carlo ensemble sampler \texttt{emcee} \citep{Foreman-Mackey:2013aa}\footnote{\protect\url{http://dan.iel.fm/emcee}}. We assume non-informative flat priors for $k$ and $\Delta z$. 


%% file: Data.tex
\section{Simulated Data}
\label{datasets}

\subsection{Buzzard simulations}

We test our calibration procedure on the \textsc{Buzzard-v1.1} simulation, a mock DES Y1 survey created from a set of dark-matter-only simulations. The simulation and creation of the mock survey data is detailed in \cite{DeRose2017, Wechsler2017, simspaper}, so we provide only a brief summary of both. \textsc{Buzzard-v1.1} is constructed from a set of 3 N-body simulations run using \textsc{L-GADGET2}, a version of \textsc{GADGET2} \citep{springel2005} modified for memory efficiency. The simulation boxes ranged from $1-4$ Gpc/h.  Light-cones from each box were constructed on the fly.  Halos were identified using \textsc{ROCKSTAR} \citep{Behroozi2013}, and galaxies were added to the simulations using the Adding Density Dependent GAlaxies to Light-cone Simulations algorithm (\textsc{ADDGALS}, \citealt{Wechsler2017}).  \textsc{ADDGALS} uses the large scale dark matter density field to place galaxies in the simulation based on the probabilistic relation between density and galaxy magnitude.  The latter is calibrated from sub-halo abundance matching in high-resolution N-body simulations. Spectral energy distributions (SEDs) are assigned to the galaxies from a training set of spectroscopic data from SDSS DR7 \citep{Cooper2011} based on local environmental density. The SEDs are integrated in the DES pass bands to generate \textit{griz} magnitudes. Galaxy sizes and ellipticities are drawn from distributions fit to deep SuprimeCam $i^{'}$-band data. The galaxy positions, shapes and magnitudes are then lensed using the multiple-plane ray-tracing code Curved-sky grAvitational
Lensing for Cosmological Light conE simulatioNS
(CALCLENS, \citealt{Becker2013}). Finally, the catalog is cut to the DES Y1 footprint, and photometric errors are added using the DES Y1 depth map \citep{Rykoff15}.

\subsection{Unknown sample in simulations - weak lensing source sample}\label{WL photoz}

\begin{figure*}
\begin{center}
\includegraphics[width=1. \textwidth]{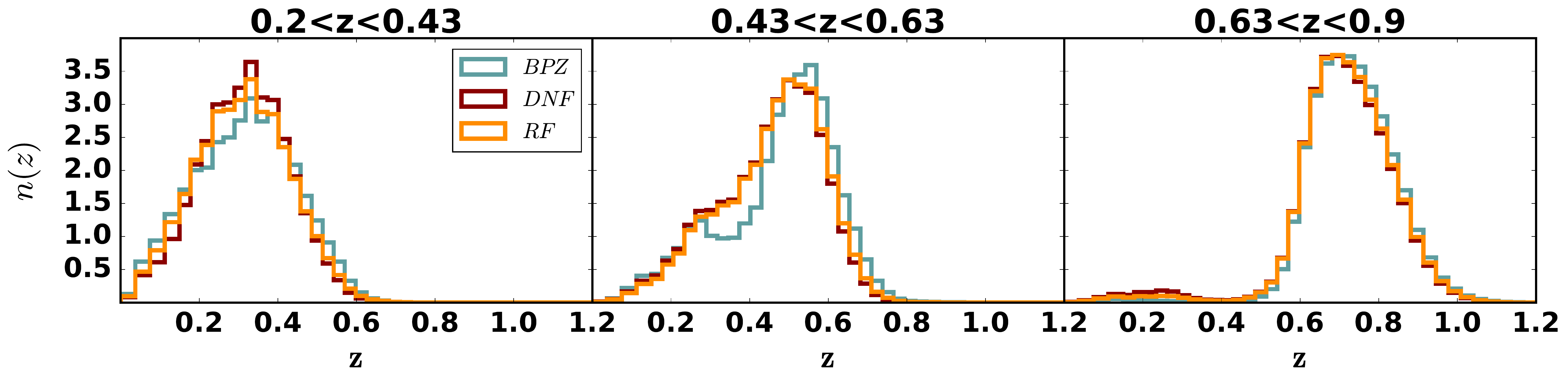}
\end{center}
\caption{True redshift distributions for the simulated weak lensing source samples obtained binning with different photo-$z$ codes, as described in $\S$\ref{WL photoz}. The redshift distributions are normalized to unity over the full redshift interval.}
\label{fig:photozposterior}
\end{figure*}

We seek to mimic the selection and redshift distribution of the weak lensing source galaxies included in the DES Y1 \texttt{METACALIBRATION} shear catalog described in \cite{shearcat}. To do so, we apply flux and size cuts to the simulated galaxies that mimic the DES Y1 source selection thresholds.  Each source has its redshift estimated, and is assigned a photometric redshift equal to the mean of the posterior redshift distribution.  These redshifts are used to divide the source galaxies into four redshift bins corresponding to [(0.2-0.43),(0.43-0.63),(0.63-0.9),(0.9-1.3)]. Due to the limited redshift coverage of the \redmagic\ reference sample, we only apply our method to the first three redshift bins. The number densities of the weak lensing sample in the simulations are $1.25, 0.82, \textrm{ and } 0.64 \/ \textrm{ arcmin}^{-2}$ for these source bins. The corresponding values of the DES Y1 shear catalog are $1.45, 1.43, \textrm{ and } 1.47 \/ \textrm{ arcmin}^{-2}$. The lower number densities in simulation have a negligible impact on the recovered statistical uncertainty, as the latter is dominated by the shot noise of the reference sample. Importantly, the shape of $\phipz(z)$ as estimated by the photo-$z$ codes in the simulations match the data with good fidelity.

Three different photo-$z$ codes have been run on the simulated WL source samples:

\begin{itemize}
\item \textbf{The Bayesian Photometric Redshifts (BPZ)} \citep{Benitez2000,Coe2006}. BPZ is a template-based method.  It returns the full probability distribution $p(z)$ for each galaxy given its magnitudes and template libraries.


\item \textbf{Directional Neighborhood Fitting (DNF)} \citep{DeVicente2016}.
DNF is a machine learning algorithm for galaxy photometric redshift estimation. Based on a training sample, DNF constructs the prediction hyperplane that best fits the neighborhood of each target galaxy in multiband flux space. It then uses this hyperplane to predict the redshift of the target galaxy, which is used to divide the sample into bins. The key feature of DNF is the definition of a new neighborhood, the Directional Neighborhood. Under this definition two galaxies are neighbors not only when they are close in the Euclidean multiband flux space, but also when they have similar relative flux in different bands, i.e. colors. 
A random sample from the photo-$z$ posterior of an object is approximated in the DNF method by the redshift of the nearest neighbor within the training sample, and it is used for the  $\phipz(z)$ reconstruction.


\item \textbf{Random Forest (RF)} \citep[e.g.][]{Breiman2001, tpz, Rau2015} RF is a Machine Learning method that generates an ensemble of decision trees from bootstrap realizations of the training data. When a new galaxy is queried down each tree in the ensemble, the decision trees vote on the galaxy redshift. The final prediction of the Random Forest is generated from the average of all decision trees' results. Both a mean redshift and full probability distribution $p(z)$ is generated for each galaxy.


\end{itemize}

Both DNF and RF photo-$z$ require us to define a training/validation sample. The sample is first defined in data. A catalog is built collecting high quality spectra from more than 30 spectroscopic surveys overlapping the DES Y1 footprint and matching them to DES Y1 galaxies \citep{photoz,Gschwend}. This catalog is then used to define the training/validation sample in simulations, by selecting the nearest neighbors in magnitude and redshift space. The selection algorithm is applied in HEALPix pixels with resolution Nside=128 (0.2 deg$^2$), so as to mimic the geometry and selection effects of the spectroscopic surveys.

The true redshift distributions of the sources binned according to each of the three photo-$z$ codes are presented in Figure~\ref{fig:photozposterior}. In what follow, we will show quantitative results for BPZ and only for one of the two machine learning codes, namely DNF, as the RF forest code does not provide results significantly different from DNF.

\subsection{Reference sample in simulations - \redmagic\ galaxies}
\label{subsec:RMG}
\begin{figure}
\centering
\includegraphics[width=.45 \textwidth]{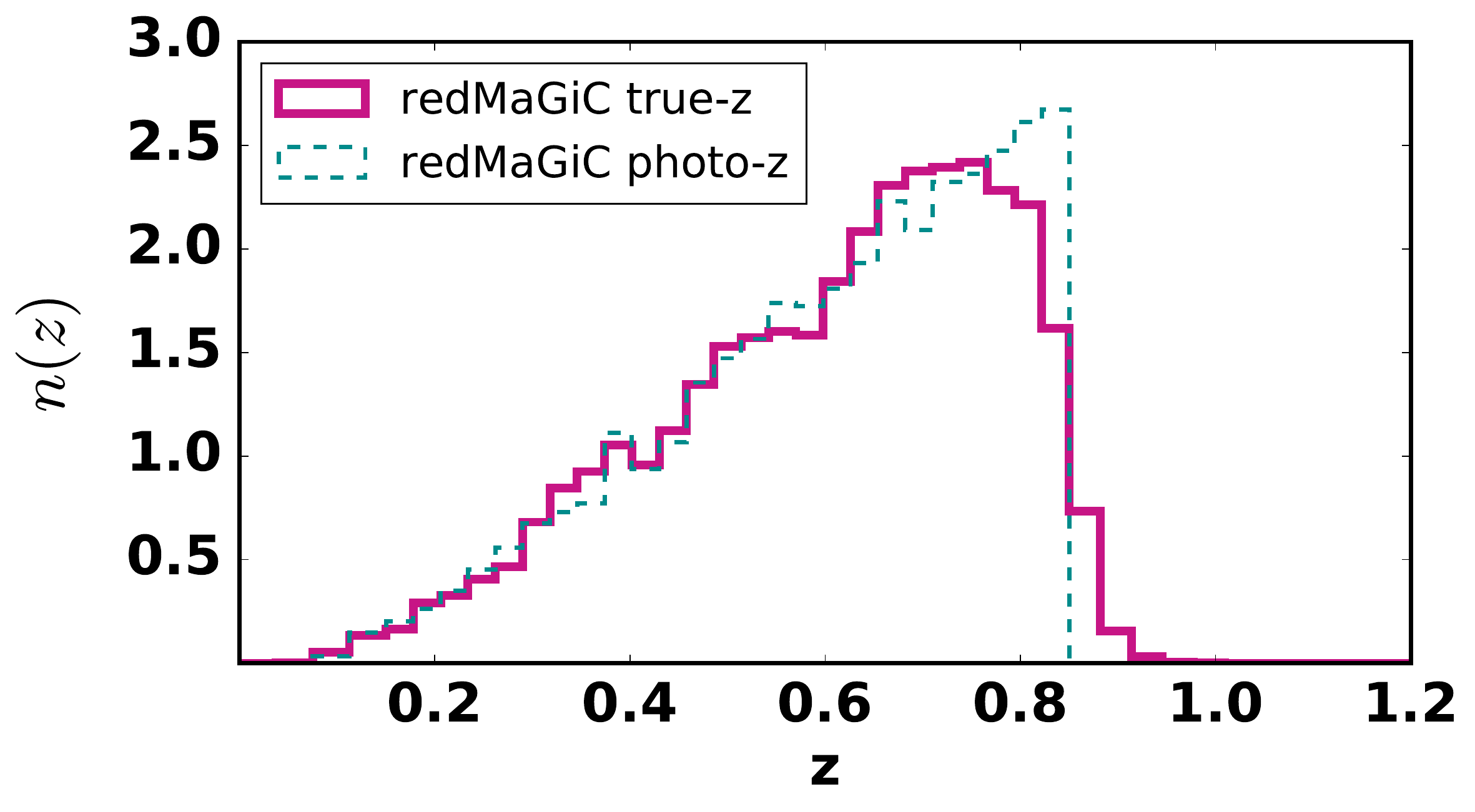}
\caption{The redshift distribution of the simulated \redmagic\ reference sample used in our analysis to measure cross-correlation redshifts. We show the distributions of both \redmagic\ true-$z$ and photo-$z$.}
\label{fig:RMGdndz}
\end{figure}

\begin{figure}
\centering
\includegraphics[width=.45 \textwidth]{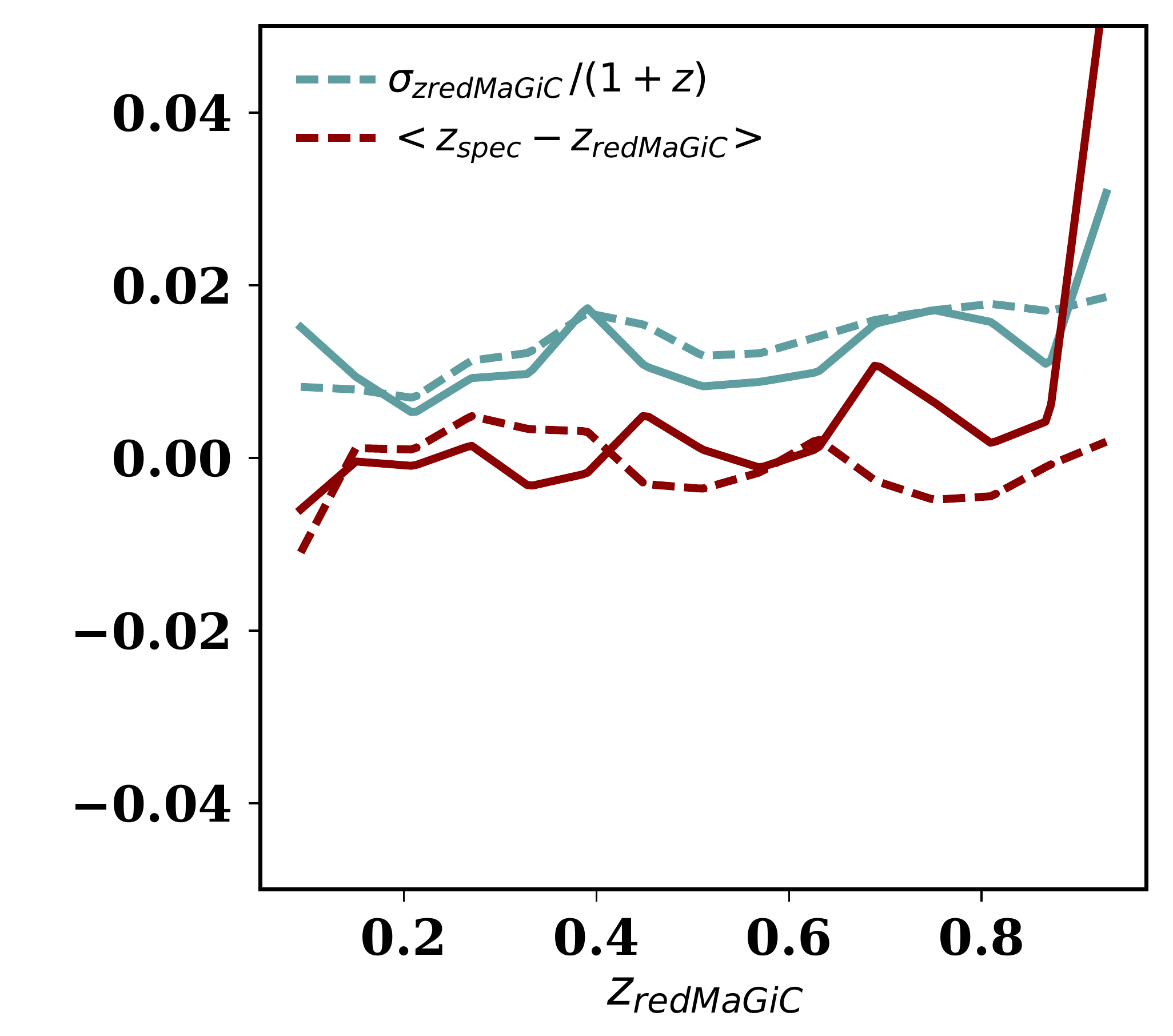}
\caption{The scatter and bias of $z_{\redmagic}$ for the simulated redmagic sample (dashed lines) compared to the data (solid lines).}
\label{fig:zred_error}
\end{figure}

We use \redmagic\ galaxies for our reference samples.  These are luminous red galaxies selected as described in \citep{Rozo2016}. The \redmagic\ algorithm is designed to select galaxies with high quality photometric redshift estimates.  This is achieved by using the red-sequence model that is iteratively self-trained by the \redmapper\ cluster finding algorithm \citep{Rykoff2014}. \textit{redMaGiC} imposes strict color cuts around this model to produce a luminosity-thresholded galaxy sample of constant comoving density.  The algorithm has only two free parameters: the desired comoving density of the sample, and the minimum luminosity of the selected galaxies. The result is a pure sample  of red-sequence galaxies with nearly Gaussian photometric redshift estimates that are both accurate and precise.

For this work we selected \redmagic\ galaxies in the redshift interval 0.15<$z$<0.85, applying the luminosity cut of $L>1.5 L_*$; the resulting redshift distribution is shown in fig. \ref{fig:RMGdndz}. The reference sample is further split into 25 uniform redshift bins. In our simulation, the mask of the \redmagic\ sample  includes all the survey regions that reach sufficient depth to render the sample volume limited up to $z$ = 0.85. Due to small differences in the evolution of the red-sequence between the simulation and the data, the simulated \redmagic\ sample has $\sim30\%$ less galaxies than the data, reaching a maximum redshift of $\zmax=0.85$ (instead of $\zmax=0.9$). We expect statistical errors in this work to be overestimated by $\sim20\%$ with respect to data\footnote{As our methodology is systematic dominated, this has a negligible impact on the results drawn in this paper.}. We note that to be consistent with the redshift interval considered here, the analysis in data has been performed cutting the \redmagic\ sample at $\zmax=0.85$ \citep{xcorr}.

The characteristic scatter and bias of the \redmagic\ photometric redshifts found in the data are very closely reproduced by the simulations as can be seen in Figure \ref{fig:zred_error}. It should be noted that in the simulation we have the true redshifts of all \redmagic\ galaxies, and thus can calculate the aforementioned statistics using the full sample, whereas in the data we only have an incomplete spectroscopic training set with which to make these measurements. \citet{redmagicpz} discusses further validation of the robustness of these estimates in the data.

We also generate a catalog of random points for \redmagic\ galaxies.  \textit{redMaGiC} randoms are generated uniformly over the footprint, as observational systematics (e.g. airmass, seeing) are not included in the simulation and for the simulated \redmagic\ sample used in this analysis, number density does not correlate with variation in the limiting magnitude of the galaxy catalogs.

%% file: Systematics_old.tex
\section{Systematic characterization}
\label{systematics}

In this section we test our clustering calibration of DES Y1 photo-$z$ redshift distributions. To assess the accuracy of the methodology, we consider the mean of the redshift distributions, computed over the full redshift interval (i.e., without restricting to the matching interval where we have reference coverage). Any residual difference in the mean between the calibrated photo-$z$ posterior and the true distribution is interpreted as a systematic uncertainty, which is quantified through the metric
\begin{equation}
\Delta \avg{z} \equiv \avg{z}_{true}-\avg{z}_{\Delta}.
\end{equation}
We will refer to this metric as the ``residual difference in the mean''. We recall that in the above equation $\avg{z}_{\Delta}$ is the mean of the photo-$z$ posterior once the photo-$z$ bias $\Delta z$ has been calibrated.

Systematic errors can arise if the clustering-based redshift distribution differs from the truth, owing to the fact that:

\begin{itemize}
\item  we are neglecting the redshift evolution of the galaxy-matter biases of both the weak lensing and \redmagic\ samples (and of the dark matter density field); hereafter we will refer to this systematic as \textbf{bias evolution systematic};
\item  we are using photo-$z$ as opposed to true redshift to bin the reference sample, hereafter referred to as  \textbf{\redmagic\ photo-$z$ systematic}.
\end{itemize}

Moreover, when we correct the photo-$z$ posterior $\phipz(z)$ using the clustering-based $\phiwz(z)$ :

\begin{itemize}
\item if the ``shape matching'' criterion is used, differences between the shapes of $\phipz(z)$ and $\phiwz(z)$  could impact the recovered photometric redshift bias, as the criterion does not impose any requirement on the mean of $\phidel(z)$ . An incorrect shape of the photo-$z$ posterior could also affect the ``mean matching'' criterion, as the matching is performed within $2 \sigma_{WZ}$ of  $\avg{z}_{\rm wz}$, and  the photo-$z$ posterior outside this interval cannot be calibrated. Hereafter we will refer to this systematic as \textbf{shape systematic}.
\end{itemize}

Below we introduce each of these systematics one at a time, computing each of their contributions to the total systematic error budget in our method. We will make the \textit{ansatz} that our systematics can be treated as independent. We will come back to this assumption later in $\S$\ref{subsect:totalsyst}.

\subsection{Bias evolution systematic} \label{subsect:bias_evolution_sys}

\begin{figure*}
\begin{center}
\includegraphics[width=1. \textwidth]{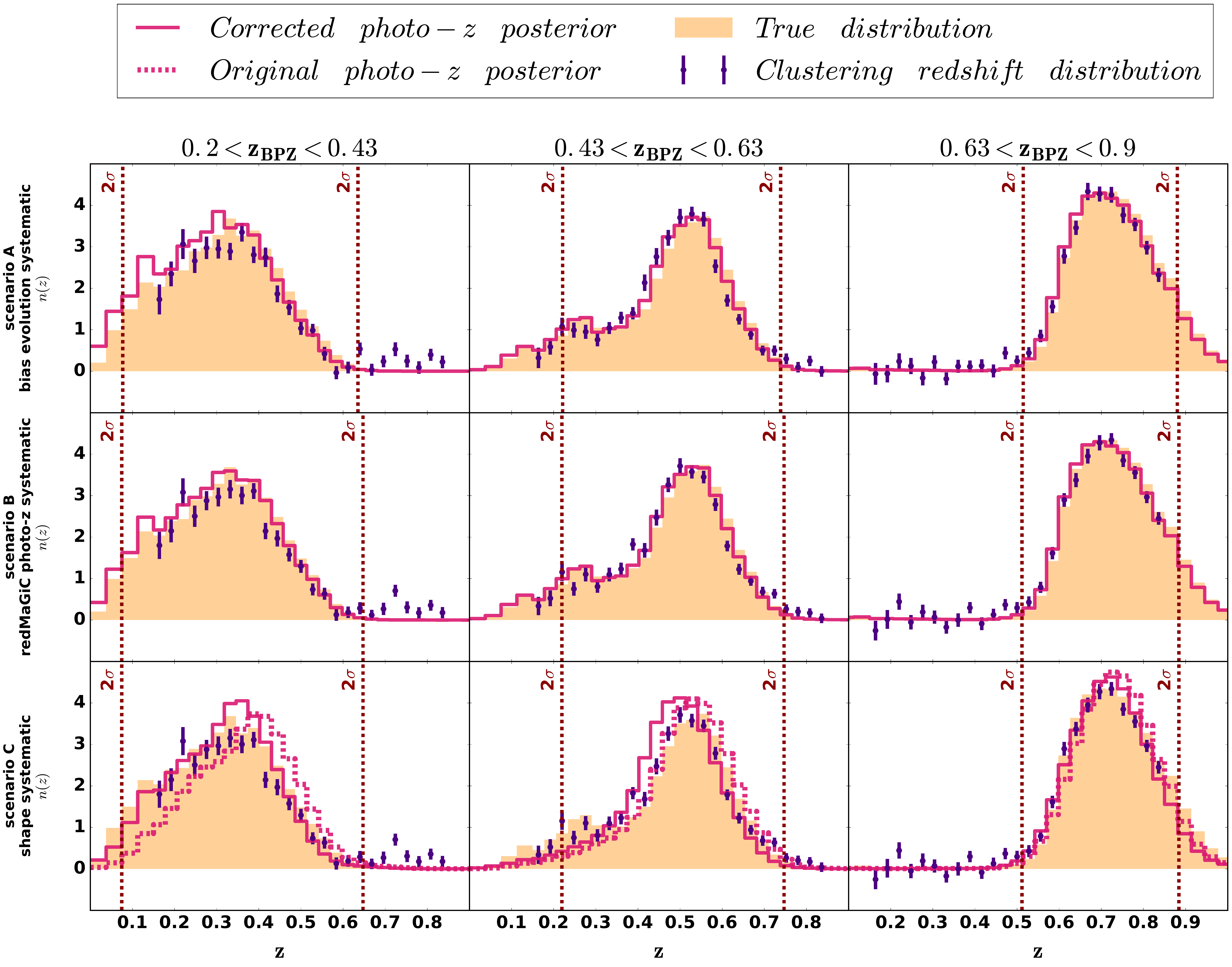}
\end{center}
\caption{Calibration procedure of the photo-$z$ posteriors of the three WL source redshift bins, for a number of different test scenarios. Points represent clustering-based estimate obtained using \redmagic\ galaxies as a reference, while the true distribution is represented by the solid yellow histogram.  The pink histograms represent the corrected photo-$z$ posterior; the uncorrected photo-$z$ posterior is shown as dashed pink histogram, only when it does not correspond to the true distribution (therefore, it is only shown for scenario C). Only the results obtained by binning with BPZ and matching the mean are shown. The $2 \sigma$ matching interval is also shown. \textit{Upper panels - scenario A: bias evolution systematic scenario}. Scenario outlined in $\S$\ref{subsect:bias_evolution_sys}; reference sample is binned using \redmagic\ true-$z$, and the true distribution is used as photo-$z$ posterior. \textit{Central panels - scenario B: \redmagic\ photo-$z$ systematic scenario}. Scenario outlined in $\S$\ref{subsect:photz_sys}; reference sample is binned using \redmagic\ photo-$z$, and the true distribution is used as photo-$z$ posterior. \textit{Lower panels - scenario C: shape systematic scenario}. Scenario outlined in $\S$\ref{subsect:shape_sys}; reference sample is binned using \redmagic\ photo-$z$, and the calibration is applied to the proper photo-$z$ posterior distribution.}
\label{fig:full_dndz}
\end{figure*} 

\begin{table*}
\caption {\textbf{BPZ systematic errors}. Systematic errors for BPZ, as a function of WL redshift bin and matching procedure. $\Delta \avg{z}_A$,$\Delta \avg{z}_B$ and $\Delta \avg{z}_C$ refers to the residual shifts in the mean relative to the scenarios A,B and C, as outlined in $\S$\ref{subsect:bias_evolution_sys}, $\S$\ref{subsect:photz_sys} and $\S$\ref{subsect:shape_sys}. For the most realistic scenario (scenario C, $\S$\ref{subsect:shape_sys}), we also show the residual shifts in the mean before and after the calibration. When a value in the table is accompanied by an uncertainty, it refers to the statistical uncertainy, estimated from the posterior of the photo-$z$ bias, as explained in $\S$\ref{correction_photoz}.}
\centering
\begin{adjustbox}{width=1.\textwidth}
\begin{tabular}{|l?c|c?c|c?c|c|}
 \hline
\multirow{3}{3cm}{}& \multicolumn{2}{c|}{\textbf{Bin 1} }& \multicolumn{2}{c|}{\textbf{Bin 2} }& \multicolumn{2}{c|}{\textbf{Bin 3} }\\
\cline{2- 7}
& mean match& shape match& mean match& shape match& mean match& shape match\\
\hline
\thead{\textbf{bias evolution}\\\textbf{systematic:}\\$\Delta \avg{z}_{A}$} &$ 0.020 $ $\pm$ $  0.006 $ &$ 0.019 $ $\pm$ $  0.005 $ &$ 0.010 $ $\pm$ $  0.004 $ &$ 0.020 $ $\pm$ $  0.002 $ &$ 0.008 $ $\pm$ $  0.003 $ &$ 0.007 $ $\pm$ $  0.003 $ \\
\hline
\thead{\textbf{\redmagic\ photo-$z$}\\\textbf{systematic:}\\$\Delta \avg{z}_{B}- \Delta \avg{z}_{A}$} &$ -0.009 $&$ -0.005 $  &$ -0.001 $&$ -0.006 $  &$ -0.001 $ &$ -0.002 $  \\
\hline
\thead{\textbf{shape}\\\textbf{systematic:}\\$\Delta \avg{z}_{C}- \Delta \avg{z}_{B}$} &$ -0.011 $ &$ -0.017 $ &$ -0.012 $  &$ -0.032 $&$ 0.004 $  &$ 0.008 $ \\
\hline
\thead{{\textbf{total systematic}}\\{\textbf{error}}} &$ \textbf{0.025}$ & \textbf{0.026} &$ \textbf{0.016}$  & \textbf{0.038}   &$ \textbf{0.014}$  & \textbf{0.011}  \\
\hline
\hline
\thead{Uncalibrated photo-$z$\\posterior $\Delta \avg{z}_C$} &$-0.048 $& -0.048 &$ -0.040 $&-0.040 &$ -0.002 $ & -0.002 \\
\hline
\thead{Calibrated photo-$z$\\posterior $\Delta \avg{z}_C$} &$0.000$ $\pm$ $0.006$& $ -0.003$ $\pm$ $0.005$  &$ -0.003$ $\pm$ $0.003$& $ -0.019$ $\pm$ $0.003$ &$ 0.011$ $\pm$ $0.002$& $ 0.013$ $\pm$ $0.002$ \\
 \hline
\end{tabular}
\end{adjustbox}
\label{table:BPZ systematics}

\centering
\caption {\textbf{DNF systematic errors}. Same as table \ref{table:BPZ systematics}, but for DNF.}
\begin{adjustbox}{width=1.\textwidth}
\begin{tabular}{|l?c|c?c|c?c|c|}
\hline
\multirow{3}{3cm}{}& \multicolumn{2}{c|}{\textbf{Bin 1} }& \multicolumn{2}{c|}{\textbf{Bin 2} }& \multicolumn{2}{c|}{\textbf{Bin 3} }\\
\cline{2- 7}
& mean match& shape match& mean match& shape match& mean match& shape match\\
 \hline
\thead{\textbf{bias evolution}\\\textbf{systematic:}\\$\Delta \avg{z}_{A}$} &$ 0.007 $ $\pm$ $  0.005 $ &$ 0.008 $ $\pm$ $  0.004 $ &$ 0.012 $ $\pm$ $  0.004 $ &$ 0.010 $ $\pm$ $  0.003 $ &$ 0.010 $ $\pm$ $  0.003 $ &$ 0.003 $ $\pm$ $  0.002 $ \\
 \hline
\thead{\textbf{\redmagic\ photo-$z$}\\\textbf{systematic:}\\$\Delta \avg{z}_{B}- \Delta \avg{z}_{A}$} &$ -0.005 $ &$ -0.002 $&$ -0.006 $  &$ -0.001 $  &$ -0.003 $ &$ -0.001 $ \\
 \hline
\thead{\textbf{shape}\\\textbf{systematic:}\\$\Delta \avg{z}_{C}- \Delta \avg{z}_{B}$} &$ -0.007 $  &$ -0.011 $  &$ -0.002 $  &$ -0.032 $ &$ -0.015 $  &$ -0.024 $  \\
 \hline
\thead{\textbf{total systematic}\\\textbf{error}} &$ {\textbf{0.012}}$ & \textbf{0.013} &$ {\textbf{0.013}}$  & \textbf{0.034} &$ {\textbf{0.019}}$  & \textbf{0.025}  \\
\hline
\hline
\thead{Uncalibrated photo-$z$\\posterior $\Delta \avg{z}_C$} &$-0.032 $& -0.032 &$ -0.048 $&-0.048 &$ -0.023 $ &-0.023 \\
\hline
\thead{Calibrated photo-$z$\\posterior $\Delta \avg{z}_C$} &$-0.005$ $\pm$ $0.005$& $ -0.005$ $\pm$ $0.003$  &$ 0.004$ $\pm$ $0.004$ & $ -0.024$ $\pm$ $0.002$ &$ -0.009$ $\pm$ $0.003$& $ -0.023$ $\pm$ $0.002$\\
 \hline
\end{tabular}
\end{adjustbox}
\label{table:DNF systematics}
\end{table*}

We first estimate errors due to bias evolution and evolution in the clustering amplitude of the density field.
We apply our method to a nearly ideal scenario in which the source galaxy distribution is binned in redshift bins according to the mean of the photo-$z$ posterior as estimated by each of the photo-$z$ codes we consider.  We use the true redshifts of the \redmagic\ reference sample when applying our cross-correlation method.  We also assume that the $\phipz(z)$ of each redshift bin is identical to the true redshift distribution.

Our results are shown in the upper panels of fig. \ref{fig:full_dndz}, labeled ``scenario A'', and the residual shifts in the mean $\Delta \avg{z}_A$ after the calibration are summarized in the first row of tables \ref{table:BPZ systematics} and \ref{table:DNF systematics}. If the calibration procedure was not affected by systematic errors, we should recover residual shifts in the mean $\Delta \avg{z}_A$ compatible with zero. However, the values in tables \ref{table:BPZ systematics} and \ref{table:DNF systematics} are as large as $|\Delta \avg{z}_A|=0.02$, owing to an incorrect $\phiwz(z)$ estimate.  All the residual shifts are substantially larger than the typical statistical uncertainty of the measurement. The specific values of the shifts vary depending on the photo-$z$ code (as they  can select slightly different populations of galaxies) and redshift bins. The two matching criteria do not always lead to the same residual shifts: in our calibration procedure, matching the shapes of two different distributions is not expected to give the same photo-$z$ bias obtained by matching their means.

We demonstrate in $\S$\ref{subsect:biasev} that correcting for the redshift evolution of the biases and of the clustering amplitude of the density field accounts for the observed residual shifts $\Delta \avg{z}_A$.  This evolution can be readily estimated in our simulated data, but is difficult to account for in the real world. Therefore, the residual shifts reported in tables \ref{table:BPZ systematics} and \ref{table:DNF systematics} represent the systematic error on the photo-$z$ bias calibration due to the bias evolution systematic.

Lastly, we note that in fig. \ref{fig:full_dndz} the clustering-based estimate recovers a spurious signal (in the form of a positive tail at high redshift) for the first redshift bin, which may potentially be explained by lensing magnification effects (see $\S$\ref{subsect:lensing}). We note, however, that the shape matching procedure is quite insensitive to biases in the tails, as the photo-$z$ posterior goes to zero.  Likewise, the mean matching within $2 \sigma_{WZ}$ of $\avg{z}_{\rm wz}$ is insensitive to the tails.

\subsection{redMaGiC photo-$z$ systematic}\label{subsect:photz_sys}

Next, we relax the assumption that we have true redshifts for the reference \redmagic\ sample.  Naively, we expect that any photometric redshift biases in \redmagic\ will imprint themselves into the clustering result.  We repeat the same analysis as before, only now we use photometric rather than true redshifts for the \redmagic\ galaxies.  Since this run is affected by bias evolution, we are interested in the \it change \rm of the residual shifts $\Delta \avg{z}$ relative to that in the previous Section.

Results are shown in the central panels of fig. \ref{fig:full_dndz}, labeled ``scenario B'', while the \textit{changes} in the residual shifts ($\Delta \avg{z}_B$-$\Delta \avg{z}_A$) are summarized in the second row of tables \ref{table:BPZ systematics} and \ref{table:DNF systematics}. These changes correspond to the values of the \redmagic\ photo-$z$ systematic. Note that we do not show the statistical uncertainty for this systematic: as the residual shifts for scenario A and B are highly correlated (since they are estimated using similar data covariances), the statistical uncertainty of their difference is close to zero.

A comparison with the values obtained for the bias evolution systematic shows that \redmagic\ photo-$z$ systematic is sub-dominant.

\subsection{Shape systematic}\label{subsect:shape_sys}

Relative to the previous run, in which the photo-$z$ posterior was assumed to be the true redshift distribution of the source galaxies, we now replace the shape of the photo-$z$ posterior by the photometrically estimated $\phipz(z)$ from each of the photo-$z$ codes we consider. This constitutes our most realistic scenario, as it suffers from all three systematics identified in this paper: bias evolution, \redmagic\ photo-$z$, and shape systematic.  Our results are shown in the lower panels of fig. \ref{fig:full_dndz}, labeled ``scenario C''. To disentangle the shape systematic from the other two, we compute the \it change \rm of the residual shift in the mean $\Delta \avg{z}$ relative to that obtained in the previous section.  The \it changes \rm in the residual shifts ($\Delta \avg{z}_C$-$\Delta \avg{z}_B$) are summarized in the third row of tables \ref{table:BPZ systematics} and \ref{table:DNF systematics}. As for the case of \redmagic\ photo-$z$ systematic, we do not show the statistical uncertainty, which should be close to zero.

We see from  tables \ref{table:BPZ systematics} and \ref{table:DNF systematics} that the shape systematic has a much stronger impact on the shape matching criteria than on the mean matching criteria. This is particularly evident in the second redshift bin, where the differences in the shapes of the photo-$z$ posterior and the true/cross-correlation redshift distributions are especially pronounced.  Note in particular the absence of a secondary low redshift peak in the photo-$z$ posteriors. Given the smaller systematic uncertainty associated with shape systematics in the mean matching criteria, we adopt it as our fiducial matching criteria for the DES Y1 analysis.

Given that this last run (scenario C) includes all systematic uncertainties, we also report in tables \ref{table:BPZ systematics} and \ref{table:DNF systematics} (fifth and sixth rows) the residual shift in the mean $\Delta \avg{z}_C$ of the photo-$z$ posterior before and after the calibration.  Error bars only account for statistical uncertainty. In almost all the cases the calibration procedure greatly reduces the residual shifts in the mean.  In particular, for many of the bins the corrected redshift distributions are consistent with zero photometric redshift bias. We note that in the second redshift bin, while it might seem by eye (fig. \ref{fig:full_dndz}) that the calibrated $\phidel(z)$ differs from the true distribution, their means are correctly matched.

While encouraging, this is partly due to canceling systematic shifts (note that many of the shifts in tables \ref{table:BPZ systematics} and \ref{table:DNF systematics} have opposite signs): the final systematic error budget - which will be discussed in \ref{subsect:totalsyst} - does not rely on such fortuitous cancellations.

\subsection{Dependence of the mean-matching criteria on the choice of redshift interval}\label{subsect:interval choice}

\begin{figure*}
\centering
\includegraphics[width=0.8 \textwidth]{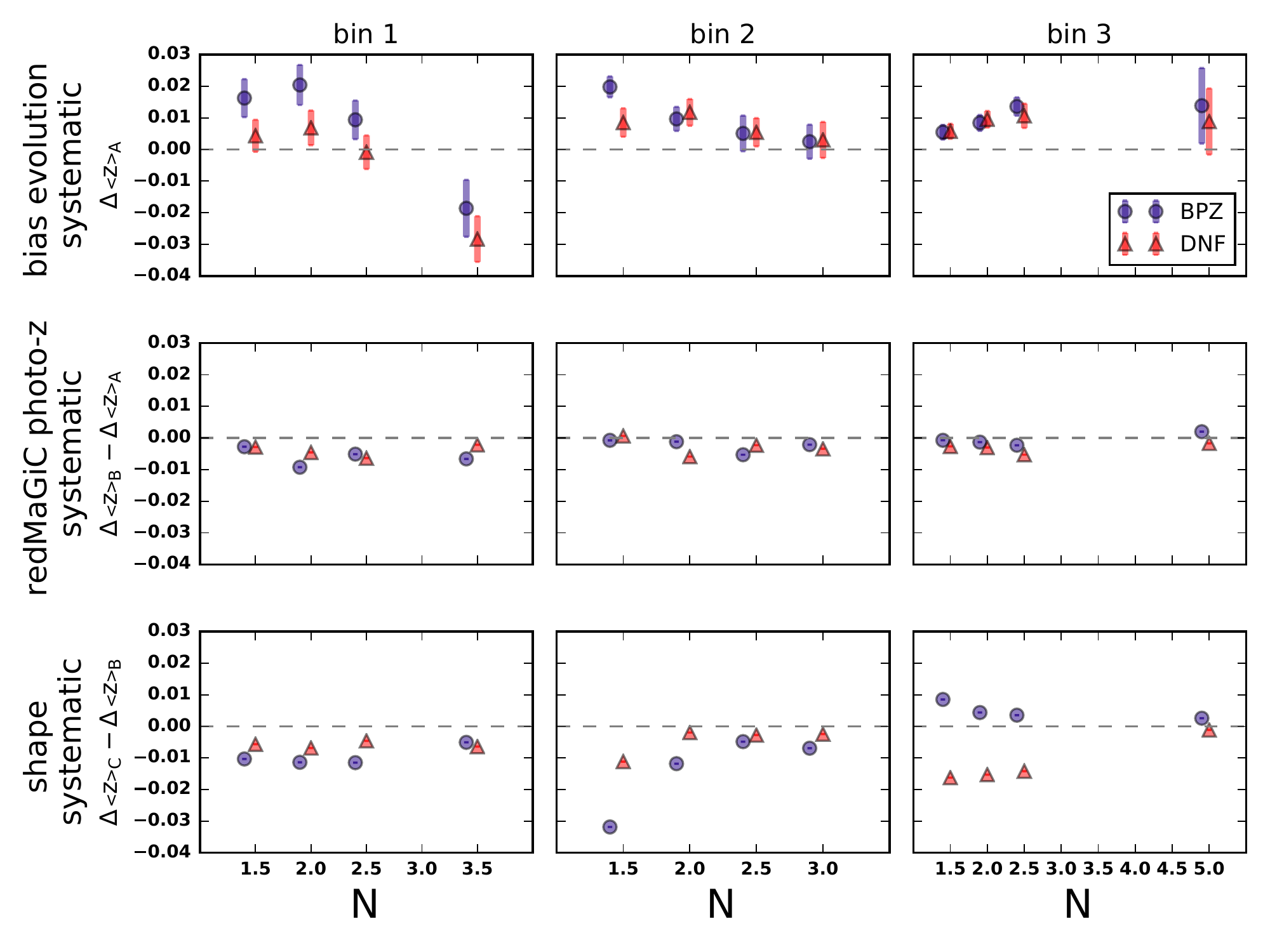}
\caption{Mean matching procedure: inferred values of the three systematics as a function of matching interval, defined as $\avg{z}_{WZ}\pm N \sigma_{WZ}$.  For each WL redshift  bin, we show results for N=1.5,N=2 and N=2.5; in addition, we show results obtained including all the redshift interval covered by \redmagic\ galaxies. This corresponds to N=3.5, N=3 and N=5 for the 1st, 2nd and 3rd source redshift bin respectively. For the 3rd bin a larger N is needed to include the low redshift tail down to z=0.15.}
\label{fig:sigmashifts}
\end{figure*}

We briefly discuss here our choice to apply the mean matching criteria only in the interval $\avg{z}_{WZ}\pm 2\sigma_{WZ}$. The interval has been chosen as it roughly  covers most of the range sampled by the true distribution, minimizing the impact of possible systematics affecting the tails of the recovered distribution.

We estimate the values of each systematic for different interval choices, namely $\avg{z}_{WZ}\pm 1.5\sigma_{WZ}$, $\pm 2\sigma_{WZ}$, $\pm 2.5\sigma_{WZ}$ as well as a run using all reference redshift bins. The results are shown in fig. \ref{fig:sigmashifts}. Variations in the values of the systematics are typically smaller than $\sim 0.005$. However, there are two exceptions. In the first WL redshift bin, large intervals ($> 2.5\sigma_{WZ}$) include in the analysis the positive tail that appears in the clustering-based estimate at high redshift. This substantially affects the bias evolution systematic. In the second WL redshift bin, the redshift interval $\avg{z_{\rm wz}} \pm 1.5\sigma_{\rm wz}$ is narrow enough that the secondary peak in the redshift distribution is not included in the analysis.  This omission introduces a larger shape systematic.

To accommodate the impact of the choice of interval in the cross-correlation measurement into our systematic error budget, for each weak lensing source bin we have opted to estimate the systematic using both our $\pm 2\sigma_{WZ}$ and $\pm 2.5\sigma_{WZ}$ cuts, always adopting the largest of the two systematic error estimates.

\subsection{Total systematic error budget}\label{subsect:totalsyst}



We choose not to correct for the biases found in $\S$\ref{subsect:bias_evolution_sys}, $\S$\ref{subsect:photz_sys} and $\S$\ref{subsect:shape_sys}, thereby not taking advantage of the fortuitous cancellations measured in the simulations. Instead, we consider each of the
shifts reported in tables \ref{table:BPZ systematics} and \ref{table:DNF systematics}  as systematic errors, and proceed to add them all up in quadrature to produce our final systematic error estimate. This assumes the three sources of systematic error to be independent. We do not expect any correlation between \redmagic\ photo-$z$ errors, and the WL galaxy-matter bias or WL photo-$z$ posterior. There might be slight (anti) correlations between the bias evolution systematic and the shape systematic, if the photo-$z$ code misplaces a population of galaxies with a given bias outside the matching interval. However, assuming a correlation coefficient of $r=0.5$ (or $r=-0.5$) between these two systematics has a negligible impact on the total error budget, so we ignored this effect.

Total systematic errors are provided in the fourth row of tables \ref{table:BPZ systematics} and \ref{table:DNF systematics}, as a function of WL source redshift bin. The total error budget is dominated by the bias evolution and shape systematic, while the \redmagic\  photo-$z$ systematic is only responsible for a marginal contribution.

Values presented in tables \ref{table:BPZ systematics} and \ref{table:DNF systematics} are substantially larger than the typical statistical uncertainty, indicating that our calibration procedure is systematic dominated. 




\subsection{Choice of method and angular scales}\label{subsect:scale_method}
\begin{figure*}
\begin{center}
\includegraphics[width=0.8 \textwidth]{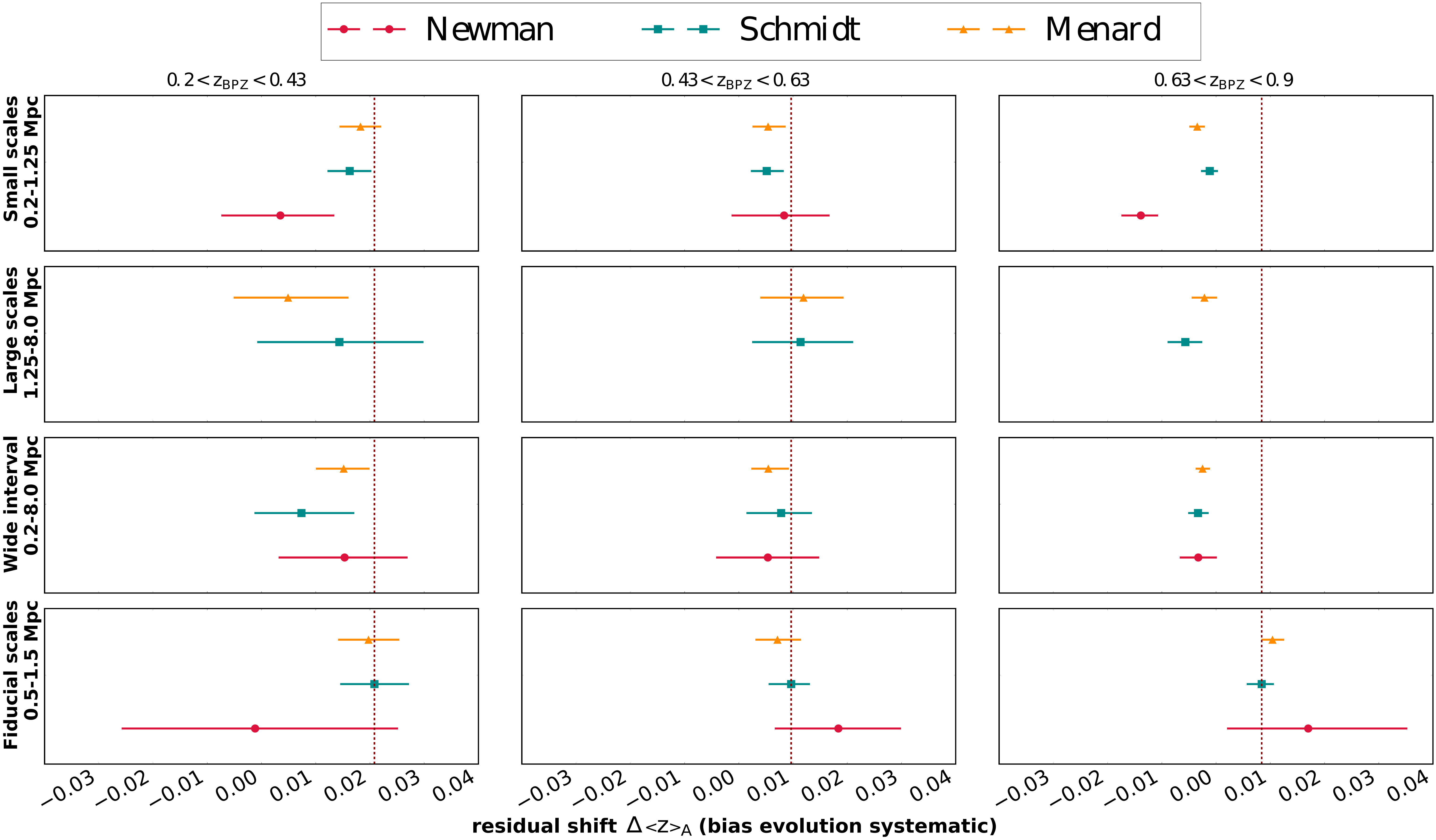}
\end{center}
\caption{Residual shifts in the mean for the scenario A outlined in $\S$\ref{subsect:bias_evolution_sys} (equivalent to the bias evolution systematic), for different choices of physical scales and different clustering-based methods. Error bars represent the statistical errors of the measurements. Vertical lines highlight our fiducial value for the bias evolution systematic, computed using the Schmidt method over scales 500kpc-1500kpc. Only the values for the mean matching procedure and for BPZ are shown. In the large scales panel, red points are missing as the Newman method failed to provide a correction.}
\label{fig:scale_method_match}
\end{figure*}

Throughout this paper we have adopted as our fiducial clustering-based method the one introduced by \cite{Schmidt2013} and considered physical scales between 500 kpc and 1500 kpc. We test the impact of the choice of angular scales by recomputing the residual shifts in the mean for one of the dominant systematic (the bias evolution systematic) with a different choice of physical scales and methods.

Figure \ref{fig:scale_method_match} shows the residual shifts using the Schmidt, M\'enard and Newman methods (outlined in $\S$\ref{method_clustering-z}), and for the following scales: 200-1250 kpc (i.e, small scales only),1250-8000 kpc (i.e, large scales only), 200-8000 kpc and 500-1500 kpc (our fiducial choice for this work).

We find that the M\'enard and Schmidt methods perform similarly. Small differences arise because of how the two methods average over angular scales. We also find that the Newman method results in the noisiest estimates, as the implementation of Newman's method introduces two new degrees of freedom that the Schmidt/M\'enard approaches do not ($\gamma_{ur}$ and $C_{ur}$, see $\S$ \ref{method_clustering-z}). For the largest angular scales (1250-8000 kpc), the reconstruction is so noisy that it fails to provide corrections to the photo-$z$ posteriors, as the MCMC chains fail to converge. The noisier estimates are due to the power-law fits: some bins can have degeneracies among power-law parameters (especially in the tails of the \Nz\ ), and in some cases the cross-correlation signal deviates from a pure power law shape.

Results are compatible within statistical errors for different choices of angular scales. Only the shifts obtained using Schmidt's and Menard's methods over our nominal scales in the third photo-$z$ bin are significantly different from the rest. However, the differences are safely within the bounds of our total systematic error. Fig. \ref{fig:scale_method_match} suggests that non linear galaxy--matter bias \citep[e.g.][]{Smith2007} seems to have a negligible impact on our methodology. At large scales, the S/N deteriorates, so we have chosen not to use the largest scales (>1500 kpc) in our fiducial analysis.

From fig.\ref{fig:scale_method_match}, it is clear that using scales as small as 200 kpc appears to improve the statistical and systematic uncertainties of our method relative to the 500 kpc inner scale radius.  However, the differences are small. We have opted to use the larger 500 kpc radius to avoid possible systematic uncertainties that may arise in the data but not in our simulation.  In particular, photometric contamination from nearby galaxies will become more important as the inner scale radius is reduced \citep[see][] {Applegate2014,Choi2016,Morrison2015,Simet2015}.  Our choice to set $r_{\rm min}=500$  is meant to safeguard our results against any such kind of neighboring biases.


%% file: Discussion.tex
\section{Discussion}\label{discussion}
\subsection{Impact of galaxy--matter biases on clustering-based methods}\label{subsect:biasev}

\begin{table*}
\centering
\caption {Residual shifts in the mean for the scenario A outlined in $\S$\ref{subsect:bias_evolution_sys} (equivalent to the bias evolution systematic),  using the estimator introduced in $\S$\ref{subsect:biasev}, which accounts for the redshift evolution of the bias and dark matter density field of both samples. We restricted the clustering-based estimate to the interval where it was possible to measure the autocorrelation function of the two samples.}
\begin{adjustbox}{width=1.\textwidth}
\begin{tabular}{|l?c|c?c|c?c|c?}
 \hline
\multirow{1}{3cm}{}& \multicolumn{2}{c|}{\textbf{Bin 1} }& \multicolumn{2}{c|}{\textbf{Bin 2} }& \multicolumn{2}{c|}{\textbf{Bin 3} }\\
\cline{2- 7}
& mean match& shape match& mean match& shape match& mean match& shape match\\
\hline
\thead{\textbf{bias evolution}\\\textbf{systematic (BPZ):}} &$ -0.004 $ $\pm$ $  0.013 $ &$ 0.001 $ $\pm$ $  0.007 $ &$ -0.007 $ $\pm$ $  0.007 $ &$ 0.004 $ $\pm$ $  0.003 $ &$ 0.002 $ $\pm$ $  0.003 $ &$ 0.001 $ $\pm$ $  0.002 $ \\
 \hline
\thead{\textbf{bias evolution}\\\textbf{systematic (DNF):}} &$ 0.000 $ $\pm$ $  0.009 $ &$ -0.001 $ $\pm$ $  0.006 $ &$ -0.002 $ $\pm$ $  0.005 $ &$ 0.002 $ $\pm$ $  0.004 $ &$ 0.004 $ $\pm$ $  0.003 $ &$ 0.002 $ $\pm$ $  0.003 $ \\
 \hline
\end{tabular}
\end{adjustbox}
\label{table:biascorr2}
\end{table*}

\begin{figure*}
\begin{center}
\includegraphics[width=1. \textwidth]{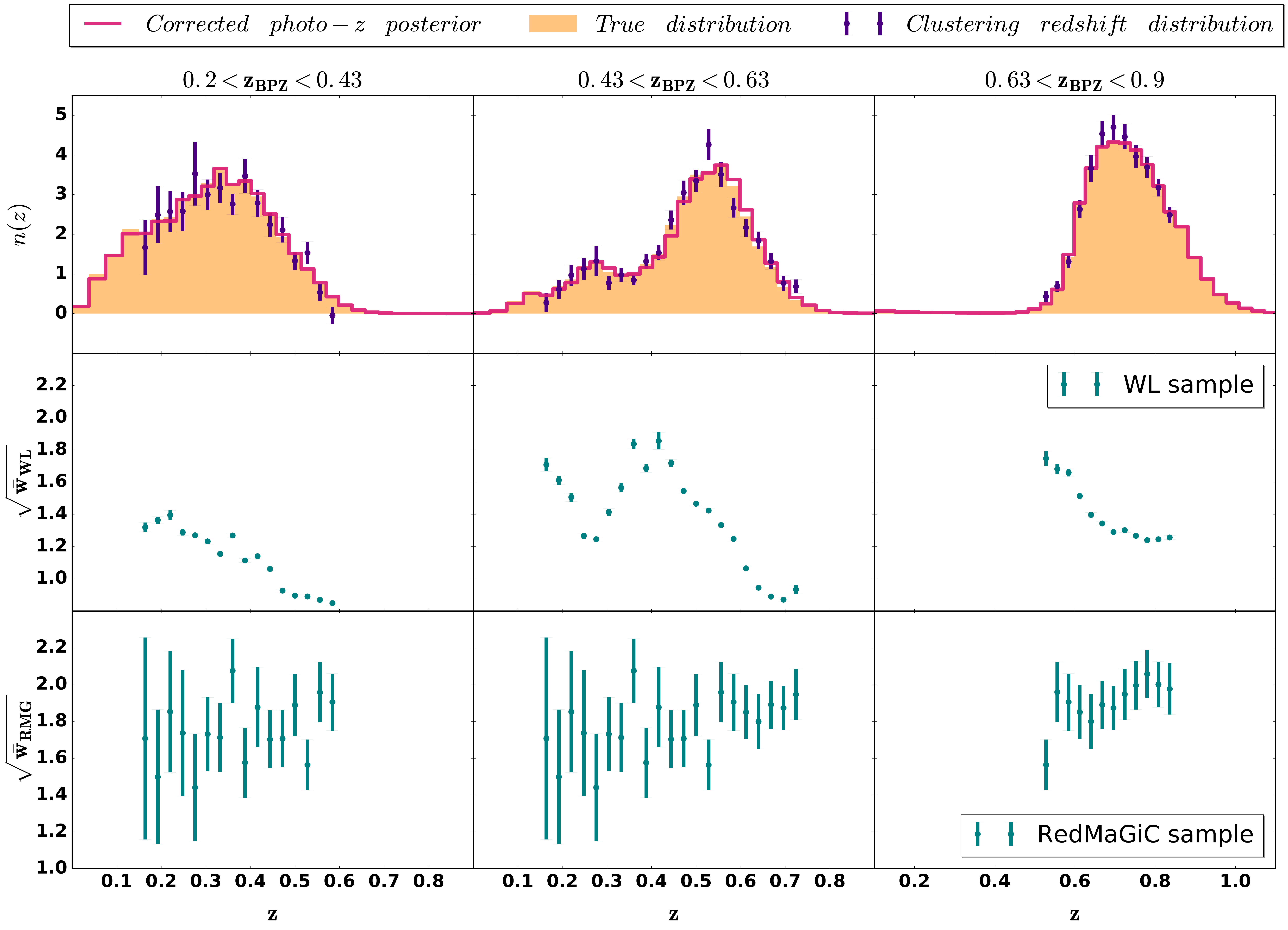}
\end{center}
\caption{WL photo-$z$ posterior calibration after correcting for the redshift evolution of the bias and dark matter density field of both samples (see $\S$\ref{subsect:biasev}). \textit{Top panels}:  clustering-based $\phiwz(z)$ and corrected photo-$z$ posteriors $\phidel(z)$. The true redshift distributions have been used as photo-$z$ posteriors.  \textit{Middle and bottom panels}: 1-bin autocorrelation estimates for the weak lensing and \redmagic\ samples, as a function of redshift. We restricted the clustering-based estimate to the interval where it was possible to measure the autocorrelation function of the two samples. Only the results obtained by binning with BPZ and matching the mean are shown. }
\label{fig:Nz_biascorr}
\end{figure*}

We now demonstrate our previous claim that the bias evolution is responsible for the systematic shifts we observed in $\S$\ref{subsect:bias_evolution_sys}, when we used true redshifts for the reference sample and the true redshift distribution as $\phipz(z)$.

In our standard methodology we have chosen not to correct for the redshift evolution of the galaxy--matter bias of both samples (and for the evolution of the dark matter density field, even if the effect is generally sub-dominant, \citealt{Menard2013}). This approximation holds as long as the biases evolve on scales larger than the typical width of the unknown distribution. In this sense, it is clear that binning the unknown sample into narrow bins through photo-$z$ or color selection helps to reduce the impact of bias evolution \citep{Menard2013,Schmidt2013,Rahman2015}. If the bins are not sufficiently narrow, neglecting the bias evolution leads to systematic shifts.

We estimate the redshift evolution of the galaxy--matter bias and dark matter density field looking at the autocorrelation functions of the reference and the unknown samples, as suggested in \cite{Menard2013} and explained in $\S$\ref{method_clustering-z}. We therefore bin both samples in 25 equally-spaced thin bins from z=0.15 to z=0.85 using their true redshift and we then measure the 1-bin autocorrelation functions of the samples. If the bins are sufficiently narrow,  and each bin has a top-hat shape, the autocorrelation functions are simply proportional to $b^2 w_{DM}$, as shown in $\S$\ref{method_clustering-z}.  We caution that we use uniform randoms to compute the WL autocorrelation functions: even though the WL sample selection function used in simulation only roughly mimics the one applied to data, using uniform random is only approximately correct.  With this caveat in mind, we estimate the redshift distribution using the following estimator:

\begin{equation}
n_u(z) \propto \frac{\bar{w}_{ur}(z)}{\sqrt{\bar{w}_{uu}(z)\bar{w}_{rr}(z)}}.
\end{equation}

The results obtained using this estimator are shown in fig. \ref{fig:Nz_biascorr} and reported in table \ref{table:biascorr2}.

By using the new estimator, we can obtain residual shifts which are compatible with zero (see the values reported in table \ref{table:biascorr2} and fig. \ref{fig:Nz_biascorr}) within statistical uncertainty. The correction induced by including in the estimator the term $\sqrt{\bar{w}_{uu}(z)}$ accounts for most of the bias evolution systematic, indicating that the major contribution to the systematic is due to the WL sample. The correction induced by the term $\sqrt{\bar{w}_{rr}(z)}$ is negligible.

We emphasize that this estimator can be implemented only in simulations, since in data we do not have access to the true redshifts needed to bin the samples. In Appendix \ref{biascorrref} we show an alternative correction obtained by binning the samples using their photo-$z$. The correction only works for the \redmagic\ bias, but we decided not to implement it as its impact is negligible.

The bias evolution of the unknown sample constitutes one of the major issues for clustering based methods, and it is one of the dominant sources of systematic in our work. It is worth noting that in our simulation the bias evolution can be complex (as it can be inferred from the middle panels of Fig. \ref{fig:Nz_biascorr}) and therefore not especially suited for correction using simple parametric approaches \citep[e.g.][]{MattewsNewman2010,Schmidt2013,DAVIS}.

As the clustering amplitudes of galaxies have been found to depend on galaxy types, colors, and luminosities \citep{Zehavi2002,Hogg2003,Blanton2006,Coil2006,Cresswell2009,marulli2013,Skibba2014}, further splitting the unknown sample into smaller subsamples with similar colors/luminosity properties (together with thinner binnings in redshift space) should alleviate the impact of bias evolution \citep{Scottez2017,vanDaalen2017}.
We also note that one could break the degeneracy between galaxy bias and redshift distribution using other probes, like galaxy-galaxy lensing \citep{gglpaper}.

\subsection{Impact of lensing magnification on clustering-based methods}

\label{subsect:lensing}

\begin{figure*}
\begin{center}
\includegraphics[width=1. \textwidth]{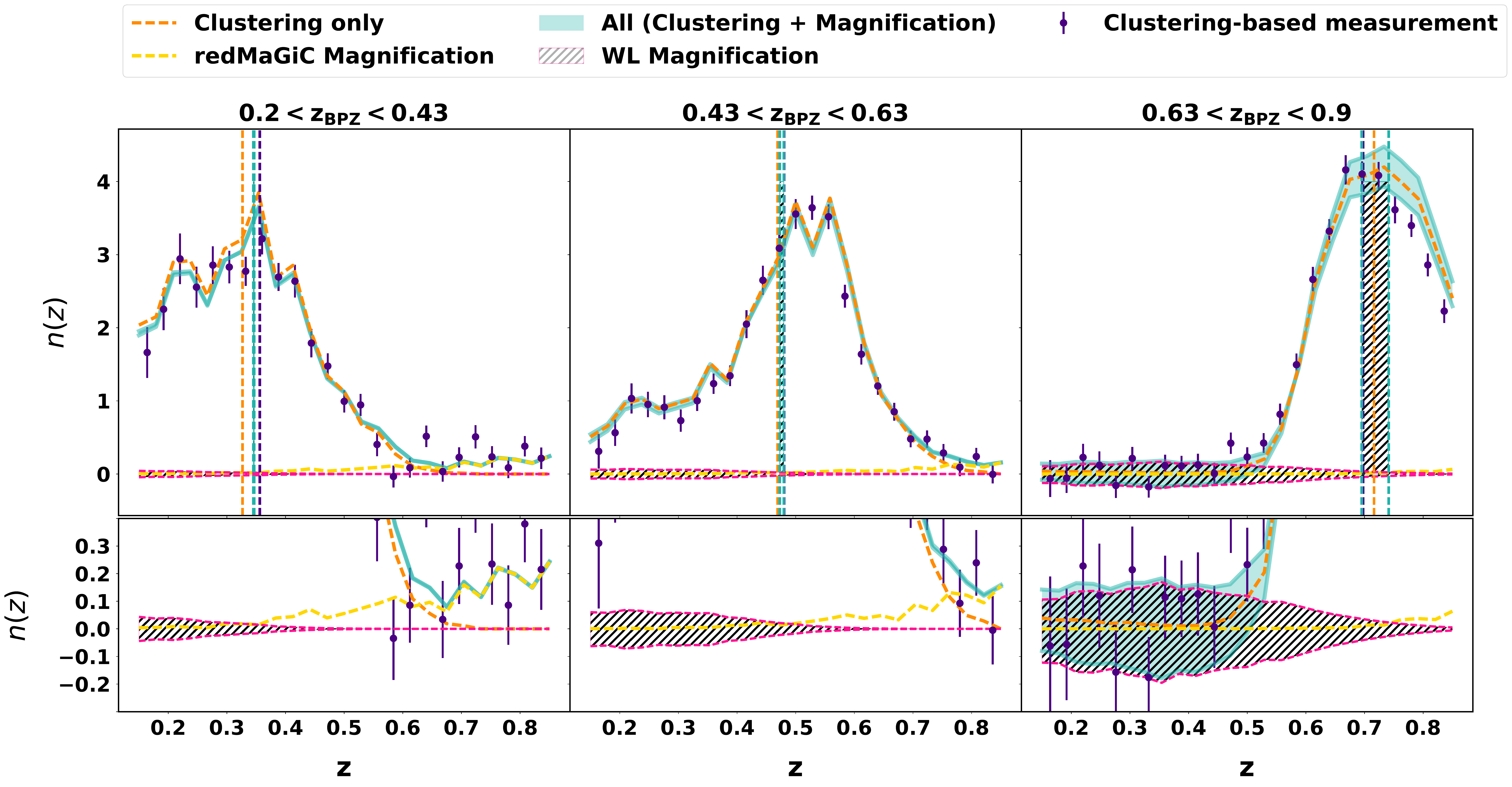}
\end{center}
\caption{\textit{Upper panels}: Effect of lensing magnification on clustering-based estimate. Shaded regions represent theoretical predictions for the various components of the signal (lensing magnification due to \redmagic\ and the WL sample, as well as the clustering due to gravitational interaction only), while the points represent the actual measurement in simulations. Vertical lines show the mean of the redshift distributions computed over the full redshift interval. \textit{Lower panels}: as the upper panels, but zoomed in on the lensing magnification signal.}
\label{fig:mag_lens}
\end{figure*}

It is well know that lensing magnification \citep{Narayan1989,Villumsen1997,Moessner1998} can lead to a change in the observed spatial density of galaxies. The enhancement in the flux of magnified galaxies can locally increase the number density, as more galaxies pass the selection cuts/detection threshold of the sample; at the same time, the same volume of space appears to cover a different solid angle on the sky, generally causing the observed number density to decrease. The net effect is driven by the slope of the luminosity function, and it has an impact on the measured clustering signal \citep{Bartelmann2001,Scranton2005,Menard2010,Morrison2012}.
For the WL samples, size bias can also be important \citep{Schmidt2009}, but it will not be considered here.

In the context of clustering-based redshift estimates, lensing magnification has been generally ignored 
(\citealt{MattewsNewman2010,Johnson2017,vanDaalen2017}, but see \citealt {McQuinn2013,Choi2016}).
\cite{Menard2013} state that the amplitude of the magnification effect on arcminute scales is generally negligible compared to the clustering signal of overlapping samples, and it has a mild dependence with redshift. However, magnification may become dominant in the regimes where the unknown and reference samples do not overlap (i.e., in the tails).

We try here to estimate the impact of lensing magnification on the recovered clustering-based \Nz . The impact of lensing magnification on the galaxy overdensity is

\begin{equation}
\delta_{obs} =\delta_{g}+\delta_{\mu},
\end{equation}
where $\delta_{g}$ is the galaxy overdensity, while $\delta_{\mu}$ is the overdensity induced by lensing magnification effects. The cross-correlation signal between two different samples is therefore

\begin{multline}
\label{eq:w_tot}
w_{ur,tot}(\theta)=  <\delta_{g,u},\delta_{g,r}>(\theta) +<\delta_{g,u},\delta_{\mu,r}>(\theta)+\\<\delta_{g,r},\delta_{\mu,u}>(\theta)+<\delta_{\mu,u},\delta_{\mu,r}>(\theta).
\end{multline}
The first term on the right side of eq. (\ref{eq:w_tot}) is associated to the clustering due to gravitational interactions, and disappears in the case of no redshift overlap between the unknown and reference samples. All of our methodology described in $\S$\ref{method_clustering-z} assumes this term to be the dominant one.
The second and third terms correspond to the lensing magnification contribution. 
The fourth term is generally small and can be neglected \citep{Duncan2014}.

Using the Limber and flat-sky approximations \citep[e.g.][] {Hui2007,Loverde2008,Choi2016}, the first clustering term in the above expression can be modeled via:

\begin{multline}
\label{eq:clustzonly}
<\delta_{g,u},\delta_{g,r}>(\theta)= b_u b_r \int \frac{dl \ l}{2\pi} J_0(l\theta)\int d\chi \\
\times  \frac{\left[n_u(z(\chi))\right]\left[n_r(z(\chi))\right]}{\chi^2H(z)} P_{NL}\left( \frac{l+1/2}{\chi},z(\chi)\right).
\end{multline}
The terms $b_u$ and $b_r$ indicate the galaxy--matter bias of the two samples; $\chi$ is the comoving distance, $H(z)$ is the Hubble expansion rate at redshift $z$. $ J_0$ is the zeroth order Bessel function. $P_{NL}(k,\chi)$ is the 3D matter power spectrum at wavenumber k (which, in the Limber approximation, is set equal to $(l + 1/2)/ \chi$) and at the cosmic time associated with redshift z.

Under the approximation of weak gravitational lensing, the terms due to lensing magnification in eq. (\ref{eq:w_tot}) can be written as


\begin{multline}
\label{eq:magonly}
 <\delta_{g,1},\delta_{\mu,2}>(\theta)= b_1 (2.5 s_2-1)  \int \frac{dl \ l}{2\pi} J_0(l\theta)\int d\chi n_1(z(\chi))\\
\times  \frac{q_2(\chi)}{\chi^2H(z)} P_{NL}\left( \frac{l+1/2}{\chi},z(\chi)\right).
  \end{multline}
The subscripts 1 and 2 are such that eq. (\ref{eq:magonly}) can  refer either to the term $<\delta_{g,u},\delta_{\mu,r}>$ or to $<\delta_{g,r},\delta_{\mu,u}>$.
The term $s_2$ is the slope of the cumulative number counts evaluated at flux limit of the sample ``2''. The slope of the cumulative number counts is formally defined for a flux limited sample as
\begin{equation}
\label{eq:slope}
s= \frac{dlog_{10} n(>m)}{dm},
\end{equation}
where $n(m)$ is the cumulative number counts as a function of magnitude $m$, and $s$ is to be evaluated at the flux limit of the sample. The term $ q_2(\chi)$ is the lensing redshift weight function defined as:
\begin{equation}
q_2(\chi) = \frac{3H_0^2\Omega_m\chi}{c^2a(\chi)} \int_{\chi}^{\chi(z=\infty)} d\chi' n_2(z(\chi'))\frac{dz}{d\chi'} \frac{\chi'-\chi}{\chi'}.
\end{equation}
$H_0$ and $a(\chi)$ are respectively the Hubble constant today and the scale factor.

Knowing the redshift distribution, the bias evolution and the slope of the cumulative number counts for the two samples, theoretical predictions for the expected clustering-based \Nz\ signal can be made through eq. (\ref{eq:clustzonly}) and eq. (\ref{eq:magonly}) and compared to the signal measured in simulations.

The true redshift distribution of the two samples is obtainable from the simulations. For the bias evolution, we make use of the 1-point estimate measured in $\S$\ref{subsect:biasev}, appropriately corrected for the contribution due to the dark matter density field. For the sake of simplicity, we do not propagate to the theoretical predictions the statistical uncertainty of the 1-point estimates of the two samples biases.

Concerning the slope of cumulative number counts, \redmagic\ galaxies are in principle not a flux-limited sample \citep[the sample is indeed volume-limited up to z=0.85, and on top of that, galaxies are required to belong to the red sequence and to have luminosity greater than a fixed threshold value, see][]{Rozo2016}. However, \redmagic\ galaxies are binned in thin redshift bins; within each bin, the sample can be well approximated as flux limited ($m>M_{lim} + 5 log_{10} (d (z_{bin}))$. The thinner the bins, the better the approximation: this should be reflected as a sharp drop in the number counts as a function of magnitude. Therefore, for each bin, we evaluate the slope of the cumulative number count using eq. (\ref{eq:slope}) at the magnitude where the number counts start to drop.

For the weak lensing sample the selection is way more complex and eq. (\ref{eq:slope}) can not be directly applied. 
Fully characterizing the selection function for the weak lensing sample goes beyond the scope of this paper. We consider the predicted lensing signal for two characteristic values of the amplitude parameter $2.5s-1$, namely $2.5s-1 = \pm 1.5$.

The results of this procedure are shown in fig. \ref{fig:mag_lens}. We see that the predicted magnification signal is qualitatively similar to the excess clustering observed in the simulations, suggesting that the excess shown at high redshift in the top-left panel of fig. \ref{fig:full_dndz} is indeed due to magnification induced by \redmagic\ galaxies at high redshift. Magnification due to the WL sample acting as a source is producing a noticeable effect only in the third bin, and the effect depends on the exact value of the amplitude parameter $2.5s-1$.

The result of this test shows that lensing magnification can have a non negligible impact on the clustering-based \Nz, mostly on the tails of the recovered distribution. It is worth stressing that the procedure presented in this paper is little affected by lensing magnification, as we cut out the tails from our analysis. We leave properly incorporating weak-lensing magnification effects into the analysis to future work.

%% file: Conclusion.tex
\section{Conclusions}
\label{conclusions}

Using numerical simulations, we characterize the performance of clustering-based calibration of the Dark Energy Survey Year 1 (DES Y1) redshift distributions. Our standard calibration procedure is divided into two steps: a first step where the redshift distribution of a given science sample is estimated using a clustering-based method; a second step where this estimated redshift distribution is used to correct for an overall photometric redshift bias in the posterior of traditional photo-$z$ algorithms. 

We use \redmagic\ galaxies as the reference sample for the clustering-based estimate. We show that our procedure could be applied in case of partial overlap in redshift space between the reference sample and the science sample. As for the science sample, we consider a simulated version of DES Y1 weak lensing source galaxies, divided in three redshift bins. We present the results for the photo-$z$ posterior of two different photo-$z$ codes (a template-based code, BPZ, and a machine learning code, DNF). The photo-$z$ codes are also used to bin the weak lensing source redshift bins, using their mean photo-$z$ redshift.

We identify and characterized in our procedure three main sources of systematic errors in our methodology:

\begin{itemize}
\item \textit{bias evolution systematic}: systematic error induced by neglecting the redshift evolution of the galaxy--matter biases of the WL and \redmagic\  samples and the evolution of the dark matter density field;
\item \textit{\redmagic\  photo-$z$ systematic}: systematic caused by not
using a spectroscopic sample as a reference;
\item \textit{shape systematic}: systematic due to an incorrect shape of the photo-$z$ posterior. This systematic is exacerbated if there is only a partial overlap between the \redmagic\ and WL samples. 
\end{itemize}

We find the bias evolution systematic (particularly, the effect due to the bias evolution of the WL sample) and shape systematic to dominate the total error budget. We also find statistical uncertainties in our procedure to be sub-dominant with respect to systematic errors. Total systematic errors for our calibration procedure, as a function of WL source redshift bin and photo-$z$ code, are provided in $\S$\ref{subsect:totalsyst}, and stand at the level of $\Delta \avg{z} \lesssim 0.02$. 

We further address the impact of changing our fiducial choices concerning the angular scales and method used for the clustering-based estimate, and discuss how our methodology could be improved. In particular, future works have to efficiently deal with the problem of the redshift evolution of the galaxy--matter bias of the science sample. This could be achieved by further splitting the science sample in luminosity/color cells. Other probes, like galaxy-galaxy lensing, could be also used to break the degeneracy between galaxy bias and redshift distribution. Lensing magnification, whose impact is marginal in this study, might no longer be negligible as survey requirements become more stringent. Lastly, we note that as clustering-based methods improve and systematic errors become sub-dominant with respect to statistical errors,  full modeling of the cross-covariance between clustering-based \Nz\ and other 2-point correlation functions will be required so as not to bias the cosmological analysis.

The calibration strategy presented in this paper is fully implemented in the DES Y1 cosmic shear and combined two-point function analysis \citep{shearcorr,keypaper}. Its direct application to DES Y1 data is discussed in two other companion papers \citep{xcorr,redmagicpz}. Even though we show systematic errors to dominate over statistical uncertainties for this calibration procedure, this does not have negative implications for the DES Y1 cosmological analysis, which remains statistically dominated.


%% file: Aknowledgements.tex
\section*{Acknowledgements}
This paper has gone through internal review by the DES collaboration. It has been assigned DES paper id DES-2017-0261 and FermiLab Preprint number PUB-17-317-A-AE.

Support for DG was provided by NASA through
Einstein Postdoctoral Fellowship grant number PF5-
160138 awarded by the Chandra X-ray Center, which
is operated by the Smithsonian Astrophysical Observatory
for NASA under contract NAS8-03060. ER acknowledges support by the DOE Early Career Program, DOE grant DE-SC0015975, and the Sloan Foundation, grant FG-2016-6443. 

Funding for the DES Projects has been provided by the U.S. Department of Energy, the U.S. National Science Foundation, the Ministry of Science and Education of Spain, 
the Science and Technology Facilities Council of the United Kingdom, the Higher Education Funding Council for England, the National Center for Supercomputing 
Applications at the University of Illinois at Urbana-Champaign, the Kavli Institute of Cosmological Physics at the University of Chicago, 
the Center for Cosmology and Astro-Particle Physics at the Ohio State University,
the Mitchell Institute for Fundamental Physics and Astronomy at Texas A\&M University, Financiadora de Estudos e Projetos, 
Funda{\c c}{\~a}o Carlos Chagas Filho de Amparo {\`a} Pesquisa do Estado do Rio de Janeiro, Conselho Nacional de Desenvolvimento Cient{\'i}fico e Tecnol{\'o}gico and 
the Minist{\'e}rio da Ci{\^e}ncia, Tecnologia e Inova{\c c}{\~a}o, the Deutsche Forschungsgemeinschaft and the Collaborating Institutions in the Dark Energy Survey. 

The Collaborating Institutions are Argonne National Laboratory, the University of California at Santa Cruz, the University of Cambridge, Centro de Investigaciones Energ{\'e}ticas, 
Medioambientales y Tecnol{\'o}gicas-Madrid, the University of Chicago, University College London, the DES-Brazil Consortium, the University of Edinburgh, 
the Eidgen{\"o}ssische Technische Hochschule (ETH) Z{\"u}rich, 
Fermi National Accelerator Laboratory, the University of Illinois at Urbana-Champaign, the Institut de Ci{\`e}ncies de l'Espai (IEEC/CSIC), 
the Institut de F{\'i}sica d'Altes Energies, Lawrence Berkeley National Laboratory, the Ludwig-Maximilians Universit{\"a}t M{\"u}nchen and the associated Excellence Cluster Universe, 
the University of Michigan, the National Optical Astronomy Observatory, the University of Nottingham, The Ohio State University, the University of Pennsylvania, the University of Portsmouth, 
SLAC National Accelerator Laboratory, Stanford University, the University of Sussex, Texas A\&M University, and the OzDES Membership Consortium.

Based in part on observations at Cerro Tololo Inter-American Observatory, National Optical Astronomy Observatory, which is operated by the Association of 
Universities for Research in Astronomy (AURA) under a cooperative agreement with the National Science Foundation.

The DES data management system is supported by the National Science Foundation under Grant Numbers AST-1138766 and AST-1536171.
The DES participants from Spanish institutions are partially supported by MINECO under grants AYA2015-71825, ESP2015-88861, FPA2015-68048, SEV-2012-0234, SEV-2016-0597, and MDM-2015-0509, 
some of which include ERDF funds from the European Union. IFAE is partially funded by the CERCA program of the Generalitat de Catalunya.
Research leading to these results has received funding from the European Research
Council under the European Union's Seventh Framework Program (FP7/2007-2013) including ERC grant agreements 240672, 291329, and 306478.
We  acknowledge support from the Australian Research Council Centre of Excellence for All-sky Astrophysics (CAASTRO), through project number CE110001020.

This manuscript has been authored by Fermi Research Alliance, LLC under Contract No. DE-AC02-07CH11359 with the U.S. Department of Energy, Office of Science, Office of High Energy Physics. The United States Government retains and the publisher, by accepting the article for publication, acknowledges that the United States Government retains a non-exclusive, paid-up, irrevocable, world-wide license to publish or reproduce the published form of this manuscript, or allow others to do so, for United States Government purposes.

This research used computing resources at SLAC National Accelerator Laboratory, and at the National Energy Research Scientific Computing Center, a DOE Office of Science User Facility supported by the Office of Science of the U.S. Department of Energy under Contract
No. DE-AC02-05CH11231.
This research was funded partially by the Australian Government through the Australian Research Council through project DP160100930.

%% file: Appendix.tex
\section{Results for a different \redmagic\ galaxy sample}
\label{AppendixA}
In this paper we have adopted \redmagic\ galaxies as a reference sample, as opposed to the more standard choice of using spectroscopic samples  \citep[e.g.][]{Menard2013,Schmidt2013,Choi2016,Hildebrandt2017}. This choice has been mainly driven by the necessity of reducing the impact of shot noise and cosmic variance that the use of a small spectroscopic sample would have implied. We proved in $\S$\ref{subsect:photz_sys} that the systematic error induced by \redmagic\ photo-$z$ is small compared to other source of systematics.

Despite statistical uncertainty being sub-dominant with respect to systematic errors, we note that the constant comoving density cut (together with luminosity threshold) used to select \redmagic\ galaxies leads to large shot noise in the lowest redshift bins. We could select the \redmagic\ sample imposing a lower luminosity threshold but a higher comoving density, so as to reduce shot noise.
We therefore create a \textit{combined} \redmagic\ galaxy sample, made of three subsamples selected as follows: 1) high density sample, 0.15<z<0.6, $L>0.5L*$; 2) high luminosity sample, 0.6<z<0.75, $L>L*$; 3) higher luminosity sample, 0.75<z<085, $L>1.5L*$. The latter corresponds to the sample used in the main analysis, but restricted to a smaller redshift interval.

We repeated the full analysis for this new \redmagic\ combined sample: results for the total systematic are summarized in table \ref{tablecombined}. As compared to our fiducial analysis (Tables \ref{table:BPZ systematics} and \ref{table:DNF systematics}), we find larger systematics for the first WL source redshift bin and slightly smaller ones for the third bin.  In general, lowering the luminosity threshold of the \redmagic\ algorithm allows to select more galaxies, but at the same time,  increases the photometric error (and the \redmagic\ photo-$z$ systematic). Moreover, being now the sample made of three subsamples each characterized by a different luminosity, we might expect a non negligible bias evolution for the reference sample. The increase in the photometric error particularly affects the first bin (as it overlaps mainly with the high density sample), and together with the stronger bias evolution, leads to a larger total systematic error. As for the third bin,  the stronger bias evolution of \redmagic\ cancels out with the bias evolution of the weak lensing sample, reducing the bias evolution systematic and the total systematic error.

Given the larger impact of \redmagic\ sample bias evolution and photometric errors on the total systematic budget, we preferred to use the higher luminosity sample for our analysis.

\begin{table}
\centering
\caption { \textbf{Total systematic error with \redmagic\ combined sample as a reference}. The table shows the total systematic error for the mean matching procedure, for the three WL source redshift bins and photo-$z$ codes.}
\begin{adjustbox}{width=0.38\textwidth}
\begin{tabular}{|l|c|c|c|}
 \hline
\multirow{3}{1cm}{}& \multicolumn{1}{c|}{\textbf{Bin 1} }& \multicolumn{1}{c|}{\textbf{Bin 2} }& \multicolumn{1}{c|}{\textbf{Bin 3} }\\
\hline
\thead{BPZ} &$ 0.037$ &$ 0.016$  &$ 0.007$ \\
 \hline
\thead{DNF} &$  0.021$ &$   0.015$  &$ 0.016$ \\
 \hline
\end{tabular}
\end{adjustbox}
\label{tablecombined}
\end{table}



\section{Correcting the redshift evolution of the galaxy--matter bias with autocorrelations when spectroscopic redshifts are not available}
\label{biascorrref}

\begin{figure*}
\begin{center}
\includegraphics[width=1 \textwidth]{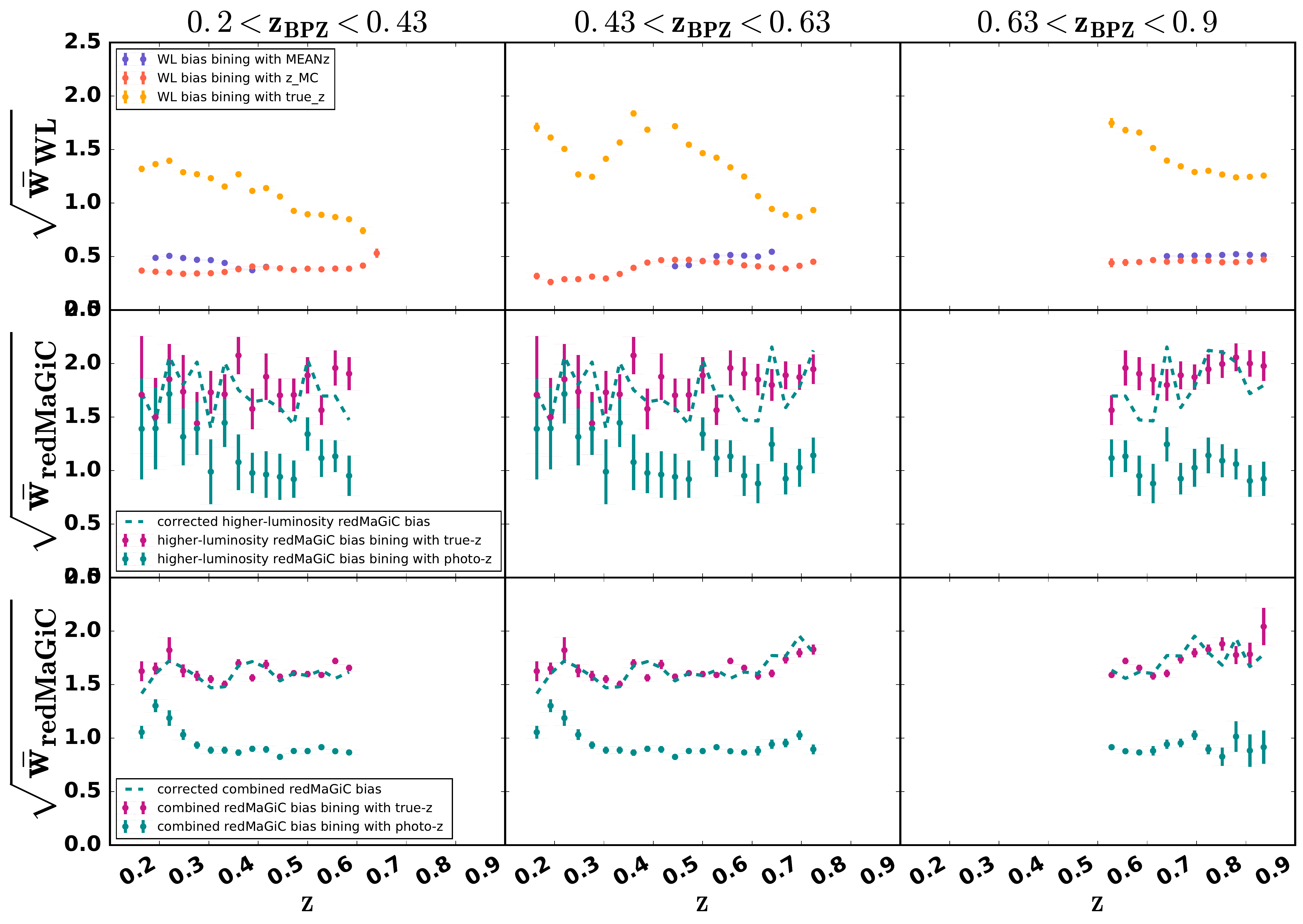}
\end{center}
\caption{\textit{Top panels} : Redshift dependence of the 1-bin estimates of the autocorrelation functions for the three WL source redshift bins. In yellow, the values obtained binning the samples using their true-$z$. In purple, the ones obtained binning the samples using their mean photo-$z$ redshift (MEANz);  in red the values obtained binning the samples using a random draw from their photo-$z$ posterior ($z_{MC}$). \textit{Middle and bottom panels}: Redshift dependence of the 1-bin estimates of the autocorrelation functions for two \redmagic\ samples - higher luminosity sample (middle panels) and combined sample (lower panels). For both samples, we display the autocorrelation functions obtained by binning the sample by true-$z$ (pink) and photo-$z$ (blue). We also display the corrected autocorrelation functions, computed starting from the estimates obtained with photo-$z$ applying the corrections explained in Appendix \ref{biascorrref}.}

\label{fig:bias_corr_WL}
\end{figure*}



In $\S$\ref{subsect:biasev} we showed that we could get rid of the bias evolution systematic within statistical errors if we could measure the autocorrelation functions of the two samples divided in thin top-hat redshift bins (i.e., using true-$z$). Unfortunately, it cannot be applied to data since we only have access to galaxies photo-$z$. Nonetheless, we could try to understand whether we could anyway correct for the redshift evolution of the galaxy--matter bias measuring the samples autocorrelation functions binned using photo-$z$.

In fig.  \ref{fig:bias_corr_WL} we show what we would obtain if we binned the WL source samples using the 1-point estimates of the photo-$z$ codes and measure the autocorrelation functions. This is compared to the results shown in $\S$\ref{subsect:biasev}, where the autocorrelation functions are binned using galaxies true-$z$ (quantities in the plot correspond to the 1-bin version of the autocorrelation functions, averaged over angular scales as explained in $\S$\ref{method_clustering-z}).

Due to the poor quality of source galaxies photo-$z$, the measurements are completely different: not only they can span a different redshift range, but also the redshift dependence is completely dissimilar.

For the reference galaxies, the scenario is a bit different (see fig. \ref{fig:bias_corr_WL}). In theory, \redmagic\ galaxies have high-quality, almost Gaussian photo-$z$s, and we could in principle try to relate the two measurements. This can be done as follows: starting from eq. (\ref{crosscorr}), the autocorrelation function included in the estimator proposed in $\S$\ref{subsect:biasev} can be written as

\begin{equation}
\label{eq11}
w_{\mathrm{RMG}}(\theta) =\int dz' b_r^2(z') n_r^2(z') w_{DM}(\theta,z'),
\end{equation}
 where  $ n_r(z')$ is the redshift distribution of the \redmagic\ galaxies in a given reference bin, and $b_r(z')$ the reference sample galaxy--matter bias. If we assume the galaxy--matter bias (and the growth factor) to evolve as a function of redshift on scales larger than the reference bin width we can re-write eq. (\ref{eq11}) as

\begin{equation}
w_{\mathrm{RMG}}(\theta) = w_{DM}(\theta,\avg{z}) b_r^2(\avg{z})\int dz'  n_r^2(z'),
\end{equation}
where the quantities outside the integral are now computed at the median redshift $\avg{z}$ of the reference bin. This would allow us to relate the 1-bin estimates of the \redmagic\ autocorrelation functions computed binning by true-$z$ and photo-$z$ as follows:

\begin{equation}
\label{eqcorr1}
\bar{w}_{\mathrm{RMG,spec-z,\avg{z}}}=\bar{w}_{\mathrm{RMG,photo-z,\avg{z}}} \frac{\int dz'  n_{\mathrm{r,spec-z}}^2(z') }{\int dz' n_{\mathrm{r,photo-z}}^2(z')}.
\end{equation}
This correction requires knowledge of
$n_{\mathrm{r,photo-z}}(z')$, which is the true distribution of the reference sample binned using \redmagic\ photo-$z$. This is usually not available in data, but an estimate can be obtained looking at the subsample of \redmagic\ galaxies with spectra.

In Fig. \ref{fig:bias_corr_WL} one can see how $\bar{w}_{\mathrm{RMG,spec-z,\avg{z}}}$ is precisely recovered using this procedure.

%% file: affiliation1.tex
$^{1}$ Institut de F\'{\i}sica d'Altes Energies (IFAE), The Barcelona Institute of Science and Technology, Campus UAB, 08193 Bellaterra (Barcelona) Spain\\
$^{2}$ Kavli Institute for Particle Astrophysics \& Cosmology, P. O. Box 2450, Stanford University, Stanford, CA 94305, USA\\
$^{3}$ Kavli Institute for Cosmological Physics, University of Chicago, Chicago, IL 60637, USA\\
$^{4}$ Universit\"ats-Sternwarte, Fakult\"at f\"ur Physik, Ludwig-Maximilians Universit\"at M\"unchen, Scheinerstr. 1, 81679 M\"unchen, Germany\\
$^{5}$ Department of Physics, Stanford University, 382 Via Pueblo Mall, Stanford, CA 94305, USA\\
$^{6}$ Centro de Investigaciones Energ\'eticas, Medioambientales y Tecnol\'ogicas (CIEMAT), Madrid, Spain\\
$^{7}$ Institute of Space Sciences, IEEC-CSIC, Campus UAB, Carrer de Can Magrans, s/n,  08193 Barcelona, Spain\\
$^{8}$ Department of Physics, University of Arizona, Tucson, AZ 85721, USA\\
$^{9}$ Instituci\'o Catalana de Recerca i Estudis Avan\c{c}ats, E-08010 Barcelona, Spain\\
$^{10}$ Department of Physics and Astronomy, University of Pennsylvania, Philadelphia, PA 19104, USA\\
$^{11}$ Laborat\'orio Interinstitucional de e-Astronomia - LIneA, Rua Gal. Jos\'e Cristino 77, Rio de Janeiro, RJ - 20921-400, Brazil\\
$^{12}$ Observat\'orio Nacional, Rua Gal. Jos\'e Cristino 77, Rio de Janeiro, RJ - 20921-400, Brazil\\
$^{13}$ SLAC National Accelerator Laboratory, Menlo Park, CA 94025, USA\\
$^{14}$ Department of Physics \& Astronomy, University College London, Gower Street, London, WC1E 6BT, UK\\
$^{15}$ Department of Physics, ETH Zurich, Wolfgang-Pauli-Strasse 16, CH-8093 Zurich, Switzerland\\
$^{16}$ Fermi National Accelerator Laboratory, P. O. Box 500, Batavia, IL 60510, USA\\
$^{17}$ Center for Cosmology and Astro-Particle Physics, The Ohio State University, Columbus, OH 43210, USA\\
$^{18}$ Department of Physics, The Ohio State University, Columbus, OH 43210, USA\\

$^{19}$ ARC Centre of Excellence for All-sky Astrophysics (CAASTRO)\\
$^{20}$ School of Mathematics and Physics, University of Queensland,  Brisbane, QLD 4072, Australia\\
$^{21}$ Centre for Astrophysics \& Supercomputing, Swinburne University of Technology, Victoria 3122, Australia\\
$^{22}$ Sydney Institute for Astronomy, School of Physics, A28, The University of Sydney, NSW 2006, Australia\\
$^{23}$ Australian Astronomical Observatory, North Ryde, NSW 2113, Australia\\
$^{24}$ The Research School of Astronomy and Astrophysics, Australian National University, ACT 2601, Australia\\
$^{25}$ Purple Mountain Observatory, Chinese Academy of Sciences, Nanjing, Jiangshu 210008, China\\
$^{26}$ Cerro Tololo Inter-American Observatory, National Optical Astronomy Observatory, Casilla 603, La Serena, Chile\\
$^{27}$ LSST, 933 North Cherry Avenue, Tucson, AZ 85721, USA\\
$^{28}$ INAF - Osservatorio Astrofisico di Torino, Pino Torinese, Italy\\
$^{29}$ Department of Astronomy, University of Illinois, 1002 W. Green Street, Urbana, IL 61801, USA\\
$^{30}$ National Center for Supercomputing Applications, 1205 West Clark St., Urbana, IL 61801, USA\\
$^{31}$ George P. and Cynthia Woods Mitchell Institute for Fundamental Physics and Astronomy, and Department of Physics and Astronomy, Texas A\&M University, College Station, TX 77843,  USA\\
$^{32}$ Department of Physics, IIT Hyderabad, Kandi, Telangana 502285, India\\
$^{33}$ Department of Physics, California Institute of Technology, Pasadena, CA 91125, USA\\
$^{34}$ Jet Propulsion Laboratory, California Institute of Technology, 4800 Oak Grove Dr., Pasadena, CA 91109, USA\\
$^{35}$ Department of Astronomy, University of Michigan, Ann Arbor, MI 48109, USA\\
$^{36}$ Department of Physics, University of Michigan, Ann Arbor, MI 48109, USA\\
$^{37}$ Instituto de Fisica Teorica UAM/CSIC, Universidad Autonoma de Madrid, 28049 Madrid, Spain\\
$^{38}$ Department of Astronomy, University of California, Berkeley,  501 Campbell Hall, Berkeley, CA 94720, USA\\
$^{39}$ Lawrence Berkeley National Laboratory, 1 Cyclotron Road, Berkeley, CA 94720, USA\\
$^{40}$ Astronomy Department, University of Washington, Box 351580, Seattle, WA 98195, USA\\
$^{41}$ Santa Cruz Institute for Particle Physics, Santa Cruz, CA 95064, USA\\
$^{42}$ Argonne National Laboratory, 9700 South Cass Avenue, Lemont, IL 60439, USA\\
$^{43}$ Departamento de F\'isica Matem\'atica, Instituto de F\'isica, Universidade de S\~ao Paulo, CP 66318, S\~ao Paulo, SP, 05314-970, Brazil\\
$^{44}$ Department of Astrophysical Sciences, Princeton University, Peyton Hall, Princeton, NJ 08544, USA\\
$^{45}$ Institute of Cosmology \& Gravitation, University of Portsmouth, Portsmouth, PO1 3FX, UK\\
$^{46}$ Brookhaven National Laboratory, Bldg 510, Upton, NY 11973, USA\\
$^{47}$ School of Physics and Astronomy, University of Southampton,  Southampton, SO17 1BJ, UK\\
$^{48}$ Instituto de F\'isica Gleb Wataghin, Universidade Estadual de Campinas, 13083-859, Campinas, SP, Brazil\\
$^{49}$ Computer Science and Mathematics Division, Oak Ridge National Laboratory, Oak Ridge, TN 37831\\
$^{50}$ Excellence Cluster Universe, Boltzmannstr.\ 2, 85748 Garching, Germany\\
$^{51}$ Max Planck Institute for Extraterrestrial Physics, Giessenbachstrasse, 85748 Garching, Germany\\